\documentclass
[prd,twocolumn,showpacs,nofootinbib,eqsecnum]{revtex4}
\usepackage{amsmath,amssymb}
\usepackage{latexsym,graphicx,epsfig}

\newcommand{\be}{\begin{equation}}
\newcommand{\ee}{\end{equation}}
\newcommand{\bes}{\begin{subequations}}
\newcommand{\ees}{\end{subequations}}
\newcommand{\bea}{\begin{eqnarray}}
\newcommand{\eea}{\end{eqnarray}}
\newcommand{\bear}{\begin{equation}\begin{array}}
\newcommand{\eear}[1]{\end{array}\label{#1}\end{equation}}
\def\ba{$$\begin{array}}
\def\ea{\end{array}$$}
\def\bra{$\begin{array}}
 \def\era{\end{array}$}

\newcommand{\bm}{\boldmath}
\newcommand{\fr}[2]{\dfrac{{ #1}}{{ #2}}}
\newcommand{\nn}{\nonumber}

\newcommand{\la}{\langle}
\newcommand{\ra}{\rangle}
\newcommand{\fn}[1]{\footnote{{\sf #1}}}
\def\vak{{\varkappa}}
\def\vep{{\varepsilon}}

\newsavebox{\fmbox}

\newenvironment{Itemize}{\begin{list}{$\bullet$} %
{\setlength{\topsep}{0.2mm}\setlength{\partopsep}{0.2mm} %
\setlength{\itemsep}{0.2mm}\setlength{\parsep}{0.2mm}}} %
{\end{list}}
\newcounter{enumct}

\newcommand{\bu}{$\bullet$\ }

\renewcommand{\Re}{\thinspace{\rm Re\thinspace}}
\renewcommand{\Im}{\thinspace{\rm Im\thinspace}}
\allowdisplaybreaks 

\begin{document}
\today
\title{Symmetries of Two Higgs Doublet Model\\ and CP violation}

\author{Ilya F. Ginzburg$^{a}$ and Maria Krawczyk$^{b}$}
\affiliation{$^a$ Sobolev Institute of Mathematics, SB RAS,
    630090 Novosibirsk, Russia}

\affiliation{$^b$ Institute of Theoretical Physics, Warsaw
University, Poland}
\begin{abstract}
We use the invariance of physical picture under a change  of
Lagrangian, the reparametrization invariance in the space of
Lagrangians and its particular case -- the rephrasing invariance,
for  analysis of the two-Higgs-doublet extension of the SM. We
found that some parameters of theory like $\tan \beta$ are
reparametrization dependent and therefore cannot be fundamental.
We use  the $Z_2$-symmetry of the Lagrangian, which prevents a
$\phi_1\leftrightarrow\phi_2$ transitions, and  the different
levels of its violation, soft and hard, to describe the physical
content of the model. In general, the  broken $Z_2$-symmetry
allows for a CP violation in the physical Higgs sector. We argue
that the 2HDM with a soft breaking of $Z_2$-symmetry is a natural
model in the description of EWSB. To simplify the  analysis we
choose among different forms of Lagrangian describing the same
physical reality a  specific one, in which the vacuum expectation
values of both Higgs fields are real.

A possible CP violation in the Higgs sector is described by
using a two-step procedure with the first step identical to
a diagonalization of the mass matrix for CP-even fields in
the CP conserving  case. We find very simple necessary
and sufficient condition for a CP violation in the Higgs
sector. We determine the range of parameters for which CP
violation and Flavor Changing Neutral Current effects are
naturally small - it corresponds to a small dimensionless
mass parameter $\nu= \Re m_{12}^2/(2v_1v_2)$. We show that
for small $\nu $ some Higgs bosons can be heavy, with mass
up to about 0.6~TeV, without violating of  the unitarity
constraints. If $\nu$ is large, all Higgs bosons except one
can be arbitrary heavy. We discuss in particular main
features of this case, which corresponds for
$\nu\rightarrow \infty $ to a decoupling of heavy Higgs
bosons.

In the  Model~II for Yukawa interactions we obtain the set of
relations among the couplings to gauge bosons and to fermions
which allows to analyse different physical situations (including
CP violation) in terms of these very  couplings, instead of the
parameters of Lagrangian.
\end{abstract}

\pacs{14.80.Cp, 12.60.Fr}

\maketitle

\section{Introduction}


A spontaneous electroweak symmetry breaking of $SU(2)\times U(1)$
(EWSB) via the Higgs mechanism is described by the Lagrangian
\begin{equation}
{ \cal L}={ \cal L}^{SM}_{ gf } +{ \cal L}_H + {\cal L}_Y \, .
\label{lagrbas}
\end{equation}
Here, ${\cal L}^{SM}_{gf}$ describes the $SU(2)\times U(1)$
Standard Model interaction of gauge bosons and fermions, ${\cal
L}_H$ is the Higgs scalar Lagrangian, and ${\cal L}_Y$ describes
the Yukawa interactions of   fermions with  Higgs scalars.

In the minimal Standard Model (SM) one scalar isodoublet with
hypercharge $Y=1$ is implemented. Here ${\cal
L}_H=(D_\mu\phi)^\dagger D_\mu\phi-V$, with  the Higgs potential
$V=\lambda\phi^4/2-m^2\phi^2/2$. A minimum of V describes the
vacuum expectation value $v$ as $\la\phi\ra= v/\sqrt{2}=
\sqrt{m^2/2\lambda}$. In this model there is one
physical Higgs boson;  its  couplings to the gauge bosons can be
expressed via masses as $g_W^{\rm SM}=\sqrt{2}M_W/v$, $g_Z^{\rm
SM}=\sqrt{2}M_Z/v$. The Yukawa interaction  has a form:
 $${\cal L}_Y =\sum
g_f^{\rm SM}\overline{Q}_L\phi q_R+h.c. \mbox{ with }\; g_f^{\rm
SM}=\sqrt{2}m_f/v.
 $$

In this paper we study in detail the simplest extension of the SM,
with one extra scalar doublet called the Two-Higgs-Doublet Model
(2HDM) which contains   more physical neutral and charged Higgs
bosons (see e.g.\ \cite{Hunter}). We treat  a CP violation in the
Higgs sector as a natural feature of the theory.

$\bullet$ This model contains two doublet fields, $\phi_1 $
and $\phi_2 $, with identical quantum numbers. Therefore,
its most general form should allow for global
transformations which mix these fields and change the
relative phase. Each such transformation generates a new
Lagrangian, with parameters given by parameters of the
incident Lagrangian and parameters of the transformation.
That is {\it the reparametrization transformation}\fn{This
very transformation is called in \cite{invappr} as Higgs
basis transformation.} of parameters of the Lagrangian.
Therefore, the physical reality described by some
Lagrangian ${\cal L}$ ({\em physical model}) is also
described by many other Lagrangians. We call this property
{\it a reparametrization invariance} in {\it a space of
Lagrangians (with coordinates given by its parameters)} and
discuss it together with its particular case -- {\it a
rephasing invariance}, in sec.~\ref{secrephas}.

If a given Lagrangian demonstrates some property, say AAA,
explicitly, we call it {\it the AAA Lagrangian or the Lagrangian of
AAA form}; a set of reparametrization equivalent Lagrangians
with the same explicit property constitutes  {\it a AAA family of
Lagrangians}.

We found that some quantities, considered often as fundamental
parameters of theory, like $\tan\beta$ -- a ratio of vacuum
expectation values of fields $\phi_1$ and $\phi_2$ -- are in fact
reparametrization dependent.

$\bullet$ One of the earliest reasons for introducing
the 2HDM was to describe the phenomenon of CP violation
\cite{Lee:1973iz}, an effect which can be potentially
large. Glashow and Weinberg \cite{Glashow:1977nt} found
that the CP violation and the flavour changing neutral
currents (FCNC) can be naturally suppressed by imposing on
the Lagrangian a $Z_2$ symmetry, that is the invariance on
the Lagrangian under the interchange
  \begin{equation}        \label{Eq:Z2-symmetry}
\phi_1 \leftrightarrow\phi_1,\phi_2 \leftrightarrow-\phi_2 \;
\text{or} \; \phi_1 \leftrightarrow-\phi_1,
\phi_2\leftrightarrow\phi_2.
 \end{equation}
This symmetry forbids the $\phi_1\leftrightarrow\phi_2$
transitions.

The most general Yukawa interaction ${\cal L}_Y$ violates this
$Z_2$ symmetry leading to the potentially large flavor--changing
neutral-current  effects. The Yukawa interaction can lead (via
loop corrections) to the CP violation even if such violation is
absent in the basic Higgs Lagrangian. Imposing specific
constraints on ${\cal L}_Y$ allows to eliminate this source of the
CP violation.

Since in Nature both the CP violation and FCNC effects are small,
we discuss separately cases of the exact $Z_2$ symmetry (then CP
is conserved) and of different levels of its violation, soft and
hard. We consider also a general renormalizability of widely
discussed forms of 2HDM Lagrangians. We analyse these problems in
sec.~\ref{seclagrZ2}.

$\bullet$ The EWSB is described by vacuum expectation values of
fields $\phi_{1,2}$ with generally different phases. This phase
difference can be eliminated by a suitable rephasing
transformation, resulting in  the {\em  Lagrangian
in a real vacuum form} (see sec.~\ref{secvac}). We use such
Lagrangian in a particular form with coefficients of the mass
(quadratic) terms in the Higgs potential expressed by coefficients
of quartic terms of potential and vacuum expectation values.

In such form of Lagrangian the real and imaginary parts of a
coefficient at the mixed quadratic term, describing a soft
violation of $Z_2$ symmetry, have different properties. The real
part can be treated as a free parameter of theory, while the
imaginary part (describing a CP violation) is constrained by the
parameters of quartic terms of Higgs Lagrangian and the vacuum
expectation values.

$\bullet$  In sec.~\ref{secphysHig} we come forward to the
observable (physical) Higgs particles. The Goldstone modes and
charged Higgs bosons $H^\pm$ are separated easily. In the
neutral sector two isotopic doublets give after EWSB one Goldstone
mode, two pure scalars (CP-even) $\eta_1$, $\eta_2$ and one CP-odd
"pseudoscalar" $A$. These three states do generally mix leading to
the physical states $h_i\; (i=1,2,3)$ without a definite CP
parity. The interaction  of these states with matter gives
observable effects of CP violation. We construct these states by a
two--step procedure, with the first step corresponding  to the
diagonalization of a partial mass squared matrix for CP--even
neural components of Higgs doublets. This leads to the states $h$
and $H$, discussed usually in a context of the CP--conserving
case. It allows us to consider a general CP nonconserving case in
terms of states $h,\,H$ and $A$, customary in the case of CP
conservation. In these terms  analyses of CP violation effects
become very transparent and some important results can be obtained
easily.

\bu  In sect.~\ref{secYuk}   the description of Yukawa couplings
is given. A most general form of Yukawa interaction violates CP
symmetry, leads to a tree-level FCNC, and breaks $Z_2$ symmetry in
a hard way (by loop corrections). A specific form of Yukawa
interaction, in which each right--handed fermion isosinglet is
coupled to only one scalar field, $\phi_1$ or $\phi_2$, guarantees
an absence of the hard violation of $Z_2$ symmetry if this
violation is absent in the proper Higgs Lagrangian ${\cal L}_H$.
With such Yukawa sector the CP violation arises only from a
structure of the Higgs Lagrangian, and FCNC effects can be
naturally small. Here we consider the well known Model II
\cite{Hunter} in the explicit form,  which is defined with
accuracy up to the rephasing transformation.

In  the investigation of  phenomenological aspects of 2HDM it is
useful to apply  {\it  relative couplings}, defined as ratios of
the couplings of each neutral Higgs boson $h_i\, (i=1,2,3)$, to
the gauge bosons $W$ or $Z$ and to the quarks or leptons
($j=W,Z,u,d,\ell...$), to the corresponding SM couplings:
 \begin{equation}
\label{Eq:gj} \chi_j^{(i)}=g_j^{(i)}/g_j^{\rm SM}\,.
 \end{equation}
As their squared values  are in principle measurable, we treat
$\chi_j^{(i)}$ themselves as measurable quantities.

We present formulae for the relative couplings describing
interactions of the observable Higgs bosons with fermions and
gauge bosons,  and than derive the set of relations among these
couplings, including obtained by us {\it pattern and linear
relations} as  well as known {\it sum rules}.  These relations are
very useful in the analyses of different physical scenarios.

In sec.~\ref{secRC}\,  we show that these relative couplings and
the relations among them are less affected by the  radiative
corrections than  the Higgs couplings themselves.

$\bullet$  Parameters of Lagrangian are constrained by positivity
(vacuum stability) and minimum constraints, discussed in
sec.~\ref{secpert}. In most cases the physical phenomena related
to the Higgs sector are described with a good accuracy by the
lowest nontrivial order  of the perturbation theory (that is the
tree approximation for the description of the Higgs sector itself
and the one--loop approximation for the Yukawa contribution to the
Higgs-boson propagators and  Higgs couplings to the photons and
gluons). This should be reliable for not too large values of
parameters of quartic terms of the Lagrangian; we consider the
relevant {\it unitarity and perturbativity constraints} in
sec.~\ref{secpert}.C. Most of above constraints were obtained in
literature for a soft violation of $Z_2$ symmetry. We discuss main
new aspects in case of the hard violation of $Z_2$ symmetry in
sec.~\ref{secperthard}.

$\bullet$ In  2HDM there is an attractive possibility  that one of
neutral Higgs bosons $h_1$ is relatively light and similar to that
in the SM while others ($h_2$, $h_3$ and $H^\pm$) are much
heavier - it is discussed in sect.~\ref{secheavy}. The studies of
2HDM are based often on an assumption of {\it decoupling} of these
heavy Higgs bosons from the known particles, i.e. effects of these
additional Higgs bosons disappear if their masses tend to
infinity.  However, such assumption is not necessary for the
description of phenomena in the presence of heavy but not
extremely heavy new particles.

For the Higgs Lagrangian in a real vacuum form the mentioned decoupling
phenomenon is governed by a singe dimensionless parameter
$\nu\propto \Re m_{12}^2$. The mass range of possible heavy Higgs
bosons, allowed by perturbativity and unitarity constraints,
depends strongly on $\nu$. For large $\nu$ the decoupling limit is
realized, i.e. the mentioned above additional Higgs bosons can be
very heavy (and almost degenerate in masses) and moreover such
additional Higgs bosons practically decouple from the lighter
particles. We analyse briefly properties of all Higgs bosons and
their interactions in this decoupling limit.

At small $\nu$  masses of $h_2$, $h_3$ and $H^\pm$ are bounded
from above by the unitarity constraints. Such Higgs bosons can be
heavy enough to avoid observation even at next generation of
colliders. Nevertheless, some non-decoupling effects can appear
for the lightest Higgs boson. We present some sets of parameters
which realize this physical picture without decoupling, still
respecting the unitarity constraints. We argue that this
non--decoupling option of 2HDM is more {\it natural}\,\, for the
weak CP violation and FCNC (in spirit of t'Hooft's concept of
naturalness \cite{'tHooft:xb}).

\bu Sec. \ref{secsum} contains our summary and discussion of
results.

\bu In the Appendix we present
triliniear and quartic couplings of physical Higgs bosons in a
general CP violating case and give the series of useful forms for
a full collection of trilinear Higgs self-couplings in the CP
conserving, soft $Z_2$ violating case.  For the case when the
Yukawa interaction is described by Model II, we express all these
trilinear couplings via the parameter $\nu$, the masses and the
relative couplings to the gauge bosons and fermions of the
physical Higgs bosons entering the corresponding vertex.

\section{Higgs Lagrangian}\label{secLagr}


To keep the value of $\rho =M_W^2/ (M_Z^2 \cos^2\theta_W)$ equal to 1
at the tree level, one assumes in 2HDM that both scalar fields
($\phi_1$ and $\phi_2$) are weak isodoublets ($T=1/2$) with
hypercharges $Y= \pm1$ \cite{Mendez:1991gp}. We use $Y=+1$ for
both of doublets  (the other choice,\linebreak[4] $Y_1=1$,
$Y_2=-1$, is  used in the MSSM; this case is also
described by equations below with a trivial change of variables).

The most general renormalizable Higgs Lagrangian can be written as
 \begin{subequations}\label{Higgslagr}\begin{equation}
   \label{Eq:Lagr-Higgs}
{\cal L}_H = T-V\,,
\end{equation}
where $T$ is the kinetic term with $D_\mu$ being the
covariant derivative containing the EW gauge fields, and
$V$ is the Higgs potential. For 2HDM we have
 \bea
 T&\begin{array}{c}= (D_{\mu} \phi_1 )^{
\dagger}(D^{\mu} \phi_1)+ (D_{\mu} \phi_2 )^{ \dagger}(D^{\mu}
\phi_2)\\[2mm] +\vak (D_{\mu} \phi_1 )^{ \dagger}(D^{\mu} \phi_2)
+\vak^*(D_{\mu} \phi_2 )^{ \dagger}(D^{\mu} \phi_1)\,,\end{array}
\label{kinterm}\\[2mm]
&\begin{array}{c} V=\dfrac{\lambda_1}{2}(\phi_1^\dagger\phi_1)^2
+\dfrac{\lambda_2}{ 2}(\phi_2^\dagger\phi_2)^2\\[2mm]
+\lambda_3(\phi_1^\dagger\phi_1) (\phi_2^\dagger\phi_2)
+\lambda_4(\phi_1^\dagger\phi_2) (\phi_2^\dagger\phi_1)
   \\[2mm]
+\dfrac{1}{2}\left[\lambda_5(\phi_1^\dagger\phi_2)^2+{\rm
h.c.}\right]\\[2mm]
+\left\{\left[\lambda_6(\phi_1^\dagger\phi_1)+\lambda_7
(\phi_2^\dagger\phi_2)\right](\phi_1^\dagger\phi_2) +{\rm
h.c.}\right\} \end{array}\label{baspot}\\[2mm]
 &\begin{array}{c}
 -\dfrac{1}{2}\left\{m_{11}^2(\phi_1^\dagger\phi_1)+m_{22}^2(\phi_2^\dagger\phi_2)
 \right.\\[2mm]
\left.+\left[m_{12}^2 (\phi_1^\dagger\phi_2)+{\rm h.c.}\right]
\right\}\,.\end{array}\label{massL}
 \eea
  \end{subequations}
The eq.~\eqref{massL} represents a mass term.  Note that
$\lambda_{1-4}$, $m_{11}^2$ and $m_{22}^2$ are real (by
hermiticity of the potential), while the $\lambda_{5-7}$,
$m_{12}^2$ and $\vak$ are in general complex parameters.
Therefore, this potential contains 14 independent
parameters while the entire Higgs Lagrangian -- 16. We will
see that CP violation in the Higgs sector, which is a
natural feature of 2HDM, can appear only if  some of these
coefficients are complex.

\subsection{Reparametrization and rephasing invariance}\label{secrephas}

\subsubsection{Reparametrization invariance}

Our model contains  two fields with identical quantum numbers.
Therefore,  it can be described both in terms of fields $\phi_k$
$(k=1,2)$,  used in Lagrangian (\ref{Higgslagr}), and in terms of
fields $\phi'_k$ obtained from $\phi_k$ by a global unitary
transformation ${\cal {\hat F}}$ of the form:
 \bea
&\begin{pmatrix}\phi_1'\\
\phi_2'\end{pmatrix}=\hat{\cal F}\begin{pmatrix}\phi_1\\
\phi_2\end{pmatrix}\,,&\label{reparam}\\[3mm]
 &\hat{\cal F}=e^{-i\rho_0}\begin{pmatrix}
\cos\theta\,e^{i\rho/2}&\sin\theta\,e^{i(\tau-\rho/2)}\\
-\sin\theta\,e^{-i(\tau-\rho/2)}&\cos\theta\,e^{-i\rho/2}
\end{pmatrix}.&\nn
 \eea

\bu {\bf\bm In the $\vak=0$ case } the transformation
(\ref{reparam}) does not change the form of kinetic term. It
induces the changes of coefficients of Lagrangian,
$\lambda_i\to\lambda'_i$ and  $m_{ij}^2\to (m')_{ij}^2$, which we
call {\it a reparametrization transformation (RPaT)}, with
  \bes\label{newlagr}\bea
&\lambda'_1 = c^2\lambda_1+s^2\lambda_2-cs\Phi-2cs\Re(\tilde
\lambda_6 + \tilde \lambda_7),&\nn
\\
&\lambda'_2=s^2\lambda_1+c^2\lambda_2 -cs\Phi+2cs\Re(\tilde
\lambda_6 + \tilde \lambda_7),&\nn
\\
&\lambda'_3= \lambda_3+cs\Phi,\quad
\lambda'_4=\lambda_4+cs\Phi,&\nn\\
&e^{2i\rho}\lambda'_5 =\lambda_5+ &\\
&  e^{i\tau}s\left[c\Phi +2is\Im\tilde\lambda_5- 2ic\Im(\tilde
\lambda_6 -\tilde \lambda_7)\right]
,&\nn\\
 &e^{i\rho} \lambda'_6
=c^2\lambda_6-s^2\lambda_7
    +\fr{e^{i\tau}}{2}cs(\lambda_1-\lambda_2
+\Psi)
,&\nn\\
&e^{i\rho}\lambda'_7=c^2\lambda_7-s^2\lambda_6
+\fr{e^{i\tau}}{2}cs(\lambda_1-\lambda_2-\Psi) ,&\nn
 \eea
 \bear{c}
{(m')}_{11}^{2}=c^2m_{11}^2+s^2m_{22}^2-2cs\mu_{12}^2,\\[2mm]
{(m')}_{22}^{2}=s^2m_{11}^2+c^2m_{22}^2+2cs \mu_{12}^2,\\[2mm]
e^{i\rho}{(m')}_{12}^{2}=m_{12}^2+\\[2mm]
e^{i\tau}\left[
 cs(m_{11}^2-m_{22}^2)-2s^2\mu^2_{12}
 \right].\end{array}
 \ee\ees
where $c=\cos\theta$, $s=\sin\theta$,
$\mu^2_{12}=\Re(m_{12}^2e^{-i\tau})$,
$\tilde{\lambda}_5=\lambda_5e^{-2i\tau}$,
$\tilde{\lambda}_{6,7}=\lambda_{6,7}e^{-i\tau}$
 and
  \ba{c}
\Phi_0=\lambda_1+\lambda_2-2(\lambda_3+\lambda_4+\Re\tilde{\lambda}_5),
\\[3mm]
\Phi=cs\Phi_0+2(c^2-s^2)
\Re(\tilde{\lambda}_6-\tilde{\lambda}_7),\\[3mm] \Psi=
(c^2-s^2)\Phi_0-8cs\Re(\tilde \lambda_6 - \tilde
\lambda_7)+2i\Im\tilde \lambda_5.
 \ea

By construction, the Lagrangian of the form (\ref{Higgslagr}) with
coefficients $\lambda_i$, $m_{ij}^2$ and that with coefficients
$\lambda'_i$, ${(m')}_{ij}^2$ given by eq.~(\ref{newlagr})
describe the same physical reality. We call this property  a {\it
reparametrization invariance}.

The set of RPaT's (\ref{newlagr}) represents the 3--parametrical
{\it reparametrization transformation group}, with {\it three
reparametrization parameters}\fn{Similar to {\it the gauge
parameter} of gauge theories.} ($\rho$, $\theta$, $\tau$), acting
in the {\it 16-dimensional space of Lagrangians} with coordinates
given by $\lambda_{1-4}$, $\Re\lambda_{5-7}$, $\Im\lambda_{5-7}$,
$m^2_{11,22}$, $\Re(m^2_{12})$, $\Im(m^2_{12})$, $\Re\vak$,
$\Im\vak\,$. The transformation $\hat{\cal{F}}$ (\ref{reparam})
represents this very group in the space of fields $\phi_i$ ({\em
the scalar basis}). It contains in addition a free parameter
$\rho_0$, which describes {\it an overall phase freedom}.

A set of physically equivalent Higgs Lagrangians, obtained from
each other by the transformations (\ref{newlagr}), forms {\it the
reparametrization equivalent space}, being a 3-dimensional
subspace of the entire space of Lagrangians -- FIG.~1. The
parameters of Lagrangian can be determined in principle only with
accuracy up to the reparametrization freedom; the different
Lagrangians within {the reparametrization equivalent space} are
physically equivalent.

\begin{figure}[h!]
\begin{center}
  \includegraphics[width=0.45\textwidth,
  height=4.5cm]{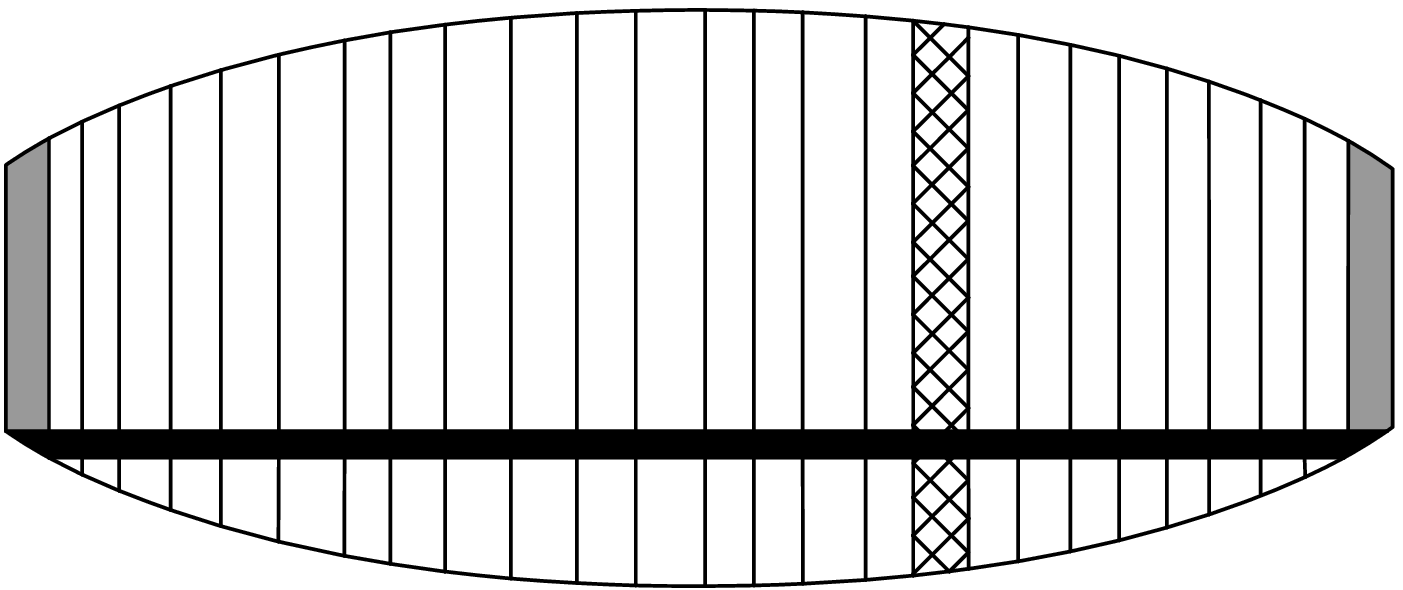}\\[3mm]
\includegraphics[width=0.45\textwidth, height=2.5cm]{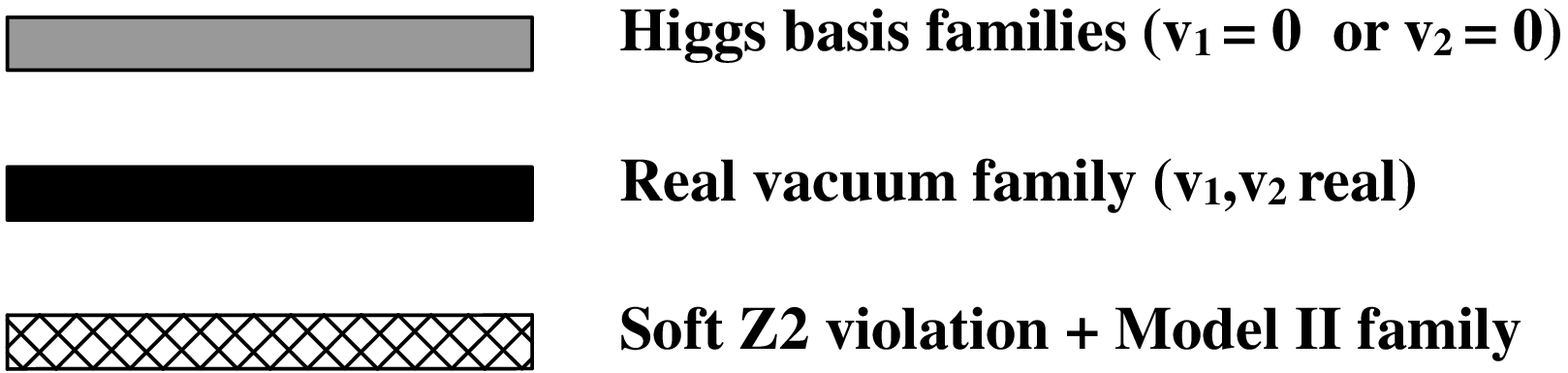}
    \caption{\it Schematic presentation of reparametrization equivalent
    space of Lagrangians. Different strips represent families
    with different  explicit properties.
A particular case when the soft $Z_2$ violating and
Model II Lagrangians families coincide is shown.  }
\end{center}
\end{figure}

All in principle observable quantities are IRpaT. That are,
for example masses of observable Higgs bosons. Each of them
is determined as eigenvalues of mass matrix
\eqref{Eq:M3by3} and \eqref{Eq:mch}. The coefficients of
secular equation for diagonalization of this mass matrix
\eqref{Eq:M3by3} (among them -- trace of this matrix and
its determinant) can be constructed from these eigenvalues.
Therefore, they are also IRpaT. The same is valid for the
eigenvalues of Higgs-Higgs scattering matrices. The set of
these IRpaT, classified in respect of isospin and
hypercharge of Higgs-Higgs system, is presented in
ref.~\cite{unitCP1}.

The  approach for construction of invariants of
reparametrization transformations (IRpaT) is proposed in
\cite{GH05}. The group--theoretical approach for
construction of all independent invariants of this
transformation is presented in ref.~\cite{Iv05}.

\bu {\bf\bm In the $\vak\neq 0$ case} the transformation
(\ref{reparam}) induces in addition change of the kinetic
term (\ref{kinterm}):
 \bea
 &T= z_1^{-1}(D_{\mu} \phi'_1 )^{
\dagger}(D^{\mu} \phi'_1)+ z_2^{-1}(D_{\mu} \phi'_2 )^{
\dagger}(D^{\mu} \phi'_2)&\nn\\
& +\vak' (D_{\mu} \phi'_1 )^{ \dagger}(D^{\mu} \phi'_2)
+\vak^{\prime*}(D_{\mu} \phi'_2 )^{ \dagger}(D^{\mu}
\phi'_1)&\label{kinetrepar}
 \eea
with
 \ba{c}
 z_1^{-1}=1-2cs\Re
(e^{-i\tau}\vak),\;\, z_2^{-1}=1+2cs\Re(e^{-i\tau}\vak)
,\\[2mm]
 \vak'=e^{-i\rho}(c^2\vak -s^2e^{2i\tau}\vak^*).\ea
So, in order to restore a canonical form of  the kinetic term a field
renormalization is needed in addition to the transformations
(\ref{newlagr}). This case will be discussed in more detail
elsewhere.\\

{\bf Remark on physical parameters.}

Some parameters of theory which are treated often as
physical (and in principle measurable) ones  are in fact
reparametrization dependent. The most important example
provides a ratio of vacuum expectation values of scalar
fields, $\tan \beta$ (\ref{veveq}). For example, under the
transformation (\ref{reparam}) with $\rho=\xi$ (see
eq.~(\ref{Eq:vevs-tanbeta})) and $\tau=0$, angle $\beta$
changes
to $\beta+\theta$.\\

\subsubsection{Rephasing invariance}

It is useful  to consider a particular case of the transformations
(\ref{reparam}) with $\theta=0$. It can be also treated as a
global transformation of fields with  independent phase rotations
({\em rephasing transformation   of the fields}):
 \bear{c}\label{rephase}
\phi_k\to e^{-i\rho_i}\phi_k,\quad (k=1,2),\\[2mm]
\rho_1=\rho_0-\fr{\rho}{2},\;\,\rho_2=\rho_0+\fr{\rho}{2},\;\,
\rho=\rho_2-\rho_1.\end{array}
 \end{equation}

This transformation leads to a change of phase of some
coefficients of Lagrangian ({\it the rephasing transformation
({RPhT})\,\, of\,\, the parameters}):
 \begin{equation}
\begin{array}{c}
\lambda_{1-4}\to \lambda_{1-4},\;\,
m_{11}^2\to m_{11}^2,\;\, m_{22}^2\to m_{22}^2,\\[3mm]
\lambda_5\to\lambda_5\, e^{-2i\rho},\;\;
\lambda_{6,7}\to\lambda_{6,7}\,e^{-i\rho},\\[2mm]m_{12}^2\to
m_{12}^2e^{-i\rho}, \,\;\vak\to \vak\,e^{-i\rho}.
\end{array}\label{Eq:rephase}
\end{equation}

By construction, the Lagrangian of the form (\ref{Higgslagr}) with
coefficients $\lambda_i$, $m_{ij}^2$ and  that with coefficients
given by eq.~(\ref{Eq:rephase}) describe the same physical
reality. We call this property  a {\it rephasing invariance}; it
is similar to the definition given in \cite{Branco}.

The transformations (\ref{Eq:rephase}) represent the
one--parametrical {\it rephasing transformation group} with
parameter $\rho$. By construction, this group is a subgroup of the
reparametrization transformation group.

The one--dimensional {\it rephasing equivalent space}, is a
subspace of the entire 3-dimensional reparametrization equivalent
space of Lagrangians. The rephasing equivalent space is given by
the sets of parameters of Lagrangians at different $\rho$. One can
say that the entire reparametrization equivalent space  is sliced
to the rephasing equivalent subspaces (represented by the
vertical strips in FIG.~1).\\

{\bf Remarks}

\bu The concept of the  rephasing invariance is easily extended to
the description of a whole system of scalars and fermions  by
adding to the transformation (\ref{Eq:rephase})
transformations (\ref{rephasingYuk})  for the Yukawa parameters.

The transformation for scalar fields (\ref{reparam})  evidently
induces changes into the set of Yukawa parameters. This may hide
some properties of the Yukawa Lagrangian, which are explicit in a
definite scalar basis (e.g. Model~I or Model~II, see
sec.~\ref{secYuk}).  The Kobyashi -- Maskawa matrix represents the
reparameterization transformation from the quark  basis of QCD  to
the electroweak basis.

\bu We will see  that  CP symmetry is conserved in the
Higgs sector if there exists a Lagrangian in the form
(\ref{Higgslagr} with all parameters real.  Obviously, this
violation does not appear if the Lagrangian with complex
parameters can be transformed by means of some RPaT
\eqref{newlagr} to a form with all parameters real.

\subsection{\bm Lagrangian and  $Z_2$ symmetry}\label{seclagrZ2}

The violation of the $Z_2$ symmetry \eqref{Eq:Z2-symmetry} in the
Lagrangian allows for the $\phi_1\leftrightarrow\phi_2$
transitions. The general Higgs Lagrangian ${\cal L}_H$
(\ref{Higgslagr}) violates $Z_2$ symmetry by terms of the operator
dimension 2 (with $m_{12}^2$), what is called {\it a soft
violation of $Z_2$ symmetry}, and of the operator dimension 4
(with $\lambda_{6,7}$ and $\vak$), called {\it a hard violation of
$Z_2$ symmetry}.

\paragraph{An exact \bm{$Z_2$} symmetry.} This case is described
by the Lagrangian ${\cal L}_H$ (\ref{Higgslagr}) with
$\lambda_6=\lambda_7=\vak= m_{12}^2 =0$ and only one parameter
$\lambda_5$ can be complex. The RPhT (\ref{Eq:rephase}) with a
suitable phase $\rho$ allows to get another form of Lagrangian
with a real $\lambda_5$, within  the rephasing invariant space.

\paragraph{A soft violation of \bm{$Z_2$} symmetry.}
In {\ the case of  soft violation of $Z_2$ symmetry} one adds to
the $Z_2$ symmetric Lagrangian the term $m_{12}^2
(\phi_1^\dagger\phi_2) + h.c. $,  with a generally complex
$m_{12}^2$ (and $\lambda_5$) parameter. This type of violation
respects the $Z_2$ symmetry at small distances (much smaller than
$1/M$) in all orders of perturbative series, i.e. the amplitudes
for $\phi_1\leftrightarrow\phi_2$ transitions disappear at
virtuality $k^2 \sim M^2 \to\infty$. That is the reason for the
name -- a "soft" violation. The  RPhT's \eqref{Eq:rephase} applied
to the Lagrangian with a softly violated $Z_2$ symmetry can
not change its character; they generate a whole {\it  soft $Z_2$
violating Lagrangian family} (the crossed "vertical" strip in
FIG.~1).

\paragraph{A hard violation of $Z_2$ symmetry.}
In the general case the terms of the operator dimension 4,  with
generally complex parameters $\lambda_6$, $\lambda_7$ and $\vak$,
are added to the Lagrangian with a softly violated $Z_2$ symmetry.
This is called a hard violation of $Z_2$ symmetry. This case
includes both the opportunity of a hidden soft $Z_2$ symmetry
violation (obtained from an exact or softly violated $Z_2$
symmetry case by a general RPaT) and of {\it the true hard
violation of $Z_2$ symmetry}, which cannot be transformed to the
case of exact or softly violated $Z_2$ symmetry by any RPaT
(\ref{newlagr}). In the latter case the $Z_2$ symmetry is broken
{\it at both large and small distances} in any scalar basis.

\subsection{The case of a hidden soft $Z_2$ violation}\label{secthid}

Let our physical system allows a description by the
Lagrangian with exact or softly violated $Z_2$ symmetry
${\cal L}_s$. The general RPaT (\ref{newlagr}) converts
this Lagrangian to a form ${\cal L}_{hs}$ with
$\lambda_6,\,\lambda_7\neq 0$ and $\vak=0$. We call ${\cal
L}_{hs}$ - a Lagrangian with a {\em hidden soft $Z_2$
violation}.

To simplify discussion of such a case  we first apply to  ${\cal
L}_s$ the RPhT (\ref{Eq:rephase}) to eliminate the phase of
$\lambda_5$. We obtain the Lagrangian ${\cal L}_s^R$ with real
$\lambda_5$ (still $m_{12}^2$ can be complex leaving open an
opportunity for CP violation). Then we apply to ${\cal L}_s^R$ a
general RPaT (\ref{newlagr})  and obtain Lagrangian ${\cal
L}_{hs}$ in  the form (\ref{Higgslagr}), with generally complex
$\lambda_5$ and nonzero $\lambda_{6,7}$  (but still $\vak=0$). We
get from (\ref{newlagr})
 \bear{c}
\lambda'_1 = c^2\lambda_1+s^2\lambda_2-cs\Phi,\\[2mm]
\lambda'_2=s^2\lambda_1+c^2\lambda_2 -cs\Phi
,\\[3mm]
\lambda'_3= \lambda_3+cs\Phi,\quad
\lambda'_4=\lambda_4+cs\Phi, \\[3mm]
\lambda'_5 =e^{-2i\rho}\lambda_5+ e^{2i\tau}[cs\Phi
+2is^2\lambda_5\sin2\tau]
,\\[2mm]
\lambda'_6 =
\fr{e^{i(\tau-\rho)}}{2}\left[cs(\lambda_1-\lambda_2)+A\right]
,\\[2mm]
\lambda'_7=
\fr{e^{i(\tau-\rho)}}{2}\left[cs(\lambda_1-\lambda_2)-A\right]
, \\[2mm]
\mbox{ with } \\[2mm]
A=(c^2-s^2)\Phi +2ics\lambda_5\sin2\tau,\\[2mm]
\Phi=cs[\lambda_1+\lambda_2-2(\lambda_3+\lambda_4+\lambda_5\cos2\tau)]\,.
\end{array}\label{softlagr}
 \ee

The eq-s (\ref{softlagr}) allow to  find parameters of the
Lagrangian ${\cal L}_s^R$ with the explicit soft violating $Z_2$
symmetry and real $\lambda_5$, once the parameters of ${\cal
L}_{hs}$
are known. The procedure is as follows:\\
 1) The value of  $\tau-\rho$ is determined from the equation
  \bes\label{hidsoftconstr}
 \be
 \fr{\lambda'_6+\lambda'_7}{\lambda^{'*}_6+\lambda^{'*}_7}= e^{2i(\tau-\rho)}\,.
 \label{1stconstr}
 \ee
2) After that one can determine  angle $\theta$ via equation
 \be
 \fr{\lambda'_6+\lambda'_7}{\lambda'_1-\lambda'_2}=e^{i(\tau-\rho)}
 \fr{\tan 2\theta}{2}\,.\label{thetdef}
 \ee
3) Next one can determine quantity $\Phi$ and
$2cs\lambda_5\sin2\tau$ via the real and imaginary parts
of
  \be
e^{-i(\tau-\rho)}(\lambda'_6\!-\!\lambda'_7)\!=\!
(c^2-s^2)\Phi\!+\!2ics\lambda_5\sin2\tau.\label{lamphidef}
  \ee
4) Then one can determine the angle $\rho$ and the parameter
$\lambda_5$ as the phase and the module of the quantity
 \be
e^{-i\rho}\lambda_5=
\lambda'_5-e^{2i(\tau-\rho)}[cs\Phi+2is^2\sin2\tau\lambda_5]\,.
\label{lamrhodef}
 \ee
 5) Finally, all remaining quantities $\lambda_{1-4}$ can be
determined easily from the first four equations (\ref{softlagr}).
\ees

Eqs. (\ref{lamphidef}) and (\ref{lamrhodef}) represent  two
different ways of obtaining the  parameter $\lambda_5$. Besides,
quantity $\Phi$ can be obtained both via eq.~(\ref{lamphidef}) and
from basic definition $\Phi=\lambda_1+\lambda_2-2[\lambda_3+
\lambda_4+ \lambda_5\cos2\tau]$. The existence of these two ways
can be considered as two constraints on the Lagrangian. It shows
explicitly that in this case the quartic sector is described by
only 8 independent parameters ($\lambda_{1-5}$ and $\theta$,
$\rho$, $\tau$) instead of 10 independent parameters of the
general Lagrangian (\ref{Higgslagr})  ($\lambda_{1-4}$,
$\Re\lambda_{5-7}$, $\Im\lambda_{5-7}$).

\subsection{Some features of the true hard $Z_2$
violation}\label{sectruehard}

\bu The most general Higgs Lagrangian \eqref{Higgslagr} cannot
be transformed to the the form with $\lambda_6=\lambda_7=0$ by any
RPaT. We denote this case as that with  {\em  true hard $Z_2$
symmetry violation}. Let us discuss briefly what should be done in
this case with the mixed kinetic terms  in Eq.~(\ref{kinterm}).
First we observe that this mixed kinetic terms can be removed by
the nonunitary transformation, e.g.
 \be\label{diagkap}
\!\!(\phi_1^{\,\prime},\!\phi_2^{\,\prime})\!\to\!\!
\left(\!\dfrac{\sqrt{\vak^*}\phi_1\!+\!\sqrt{\vak}\phi_2}
{2\sqrt{|\vak|(1\!+\!|\vak|)}}\!\pm\!
\dfrac{\sqrt{\vak^*}\phi_1\!-\!\sqrt{\vak}\phi_2}{2\sqrt{|\vak|(1\!-\!|\vak|)}}
\!\right)\!.\!\!
 \ee
However, in  presence of the $\lambda_6$ and $\lambda_7$ terms,
the renormalization of quadratically divergent, non-diagonal
two-point functions leads anyway to  the mixed kinetic terms (e.g.
from loops with $\lambda_6^*\lambda_{1,3-5}$ and
$\lambda_7^*\lambda_{2-5}$). It means that  $\vak$\ \ becomes
nonzero at the higher orders of perturbative theory, and {\it vice
versa} a  mixed kinetic term generates counter-terms with
$\lambda_{6,7}$. Therefore all of these terms should be included
in Lagrangian (\ref{Eq:Lagr-Higgs}) on the same footing, i.e. the
treatment of the hard violation of $Z_2$ symmetry without $\vak$\
terms is inconsistent (see also \cite{Ginzburg1977,Wein90}). (The
phenomenon is analogous to a need of a quartic coupling of the
form $\lambda\phi^4$ in the renormalization of the
$\bar{\psi}\gamma^5\psi\phi\;$ theory \cite{bogoliubov}.) Note
that the parameter $\vak$ is generally running like parameters
$\lambda$'s. Therefore, the Lagrangian {\it remains off--diagonal}
in fields $\phi_{1,2}$ even at very small distances, above the EWSB
transition. Such theory seems to be {\it unnatural}.

\bu To find a signature of this case in the arbitrary form of
Lagrangian  it is useful to consider a polarization operator
matrix for two fields:
${\cal P}=\begin{pmatrix}\Pi_{11}&\Pi_{12}\\
\Pi_{21}&\Pi_{22}\end{pmatrix}$. In the general case the ratio
$\Pi_{12}/(\Pi_{11}-\Pi_{22})$ is a running quantity at large
Higgs boson virtuality $k^2$ in contrast to the case of a hidden
$Z_2$ symmetry, where this ratio is not running.

Indeed, let us consider the Lagrangian with soft violation of
$Z_2$ symmetry, ${\cal L}_s$, like in sect~\ref{secthid}. The
one--loop polarization operator matrix for two fields has a form
 $
{\cal P}=\begin{pmatrix}\Pi_1^s &0\\0&\Pi_2^s\end{pmatrix}k^2
+\mbox{\it finite terms},
 $
for $k^2\to \infty$. The elements $\Pi_1^s$ and $\Pi_2^s$ describe
renormalization of fields $\phi_1$ and $\phi_2$, respectively.
There is no mixed kinetic term, and the
$\phi_1\leftrightarrow\phi_2$ transitions at small distances are
absent.

Under RPaT, the ${\cal L}_s$ is converted to the Lagrangian ${\cal
L}_{hs}$, with nonzero $\lambda_6$ and $\lambda_7$ terms
(\ref{softlagr}), still with $\vak=0$. This Lagrangian leads to
the polarization operator with nonzero mixed  term:
  \ba{c}
\fr{{\cal P}'}{k^2} = \begin{pmatrix}\Pi_1^s\cos^2\theta+\Pi_2^s\sin^2\theta&\Pi_{12}\\
\Pi_{12}^*&\Pi_2^s\cos^2\theta+\Pi_1^s\sin^2\theta\end{pmatrix}
,\\[4mm]
 \Pi_{12}=(\Pi_1^s-\Pi_2^s)e^{-i\tau}\sin\theta\cos\theta.
 \ea
Naively, this form of the polarization operator suggests that one
should introduce in the Lagrangian the mixed kinetic term
describing  transitions\linebreak[4]
$\phi'_1\leftrightarrow\phi'_2$. However, the renormalization
group analysis ensures that in this case the ratio
$\Pi_{12}/(\Pi_{11}-\Pi_{22})$ at large $k^2$ is renormalization
invariant quantity (in contrast to the mentioned above case of the
true hard violation of $Z_2$ symmetry). In such case there exist
some parameters $\rho,\,\theta,\,\tau$ which restore the incident
form of ${\cal L}_H$ with soft $Z_2$ symmetry violation, i.e.
without kinetic terms. In such scalar basis the transitions
$\phi_1\leftrightarrow\phi_2$ are absent at small distances. Since
the kinetic term of Lagrangian can be obtained from the initial
$diag(1,\,1)$ form by the orthogonal transformation
(\ref{reparam}), one can conclude that the mentioned relations
among parameters of new quartic terms prevent an appearance of the
mixed kinetic term in the Higgs Lagrangian in any
reparametrization equivalent  Lagrangians. As it was mentioned
above, this is in contrast to the general case with the true hard
violation of $Z_2$ symmetry, where $\phi_1\leftrightarrow\phi_2$
transitions at different large $k^2$ cannot be ruled out
simultaneously by any RPaT (\ref{newlagr}).

The another example is given by the EWSB procedure
(sec.~\ref{secvac}) in the case of soft violation of $Z_2$
symmetry. It  transforms the Lagrangian expressed in terms of
fields $\phi_{1,2}$ to that written in terms of Higgs fields
$h_{1-3}$ and $H^\pm$. In this form  many quartic couplings appear
but there are some relations among them, since all of them were
obtained from the initial Lagrangian ${\cal L}_s$ with 6
parameters ($\lambda_{1-4}$, $\Re\lambda_5$, $\Im\lambda_5$) and
the orthogonal transformation from the ($\phi_1,\,\phi_2$) basis
to ($H^\pm,\,h_1,\,h_2,\,h_3$) basis with the additional 3
parameters. In this Lagrangian a mixed polarization operator may
also appear, however no mixed kinetic term in contrast to the case
of true hard violation of $Z_2$ symmetry. This is due the
mentioned relations among parameters of new quartic terms which
prevent appearance of the mixed kinetic term in the Higgs
Lagrangian \cite{Pilaf97}. The detailed discussion of these
problems will be done elsewhere.

Other aspects of the hard violation of $Z_2$ symmetry are related
to the description of Yukawa sector. This will be discussed in
sec.~\ref{secYuk}.\\

{\bf Remarks}

\bu The diagonalization described  by Eq.~(\ref{diagkap}) is
rather special and it changes  even the definitions of
$\lambda$'s, what would destroy relatively simple relations
between the masses of the Higgs bosons discussed below.

\bu Although in this paper we present relations  for the case of
hard violation of $Z_2$ symmetry at $\vak=0$ one should keep in
mind that  loop corrections can change results significantly. {\it
Such treatment of the case with hard violation of $Z_2$ symmetry
is as incomplete as in most  of the  papers considering this "most
general 2HDM potential".} A full treatment of this problem goes
beyond the scope of the present paper.

\subsection{Vacuum} \label{secvac}
The extremes of the potential define the  vacuum expectation
values (v.e.v.'s $\la\phi_{1,2}\ra$) of the fields $\phi_{1,2}$ via
equations:
\begin{equation}          \label{Eq:min-cond}
\dfrac{\partial V}{\partial\phi_1}
\bigg|_{\genfrac{}{}{0pt}{}{\phi_1=\langle\phi_1\rangle,}
{\phi_2=\langle\phi_2\rangle}}=0, \qquad \dfrac{\partial
V}{\partial\phi_2}
\bigg|_{\genfrac{}{}{0pt}{}{\phi_1=\langle\phi_1\rangle,}
{\phi_2=\langle\phi_2\rangle}}=0.
\end{equation}

This equation has trivial electroweak symmetry conserving
solution $\la\phi_1\ra=0$, $\la\phi_2\ra=0$ and electroweak
symmetry violating solutions, discussed below. With accuracy to
the choice of $z$ axis in the weak isospin space, and using the
overall phase freedom of the Lagrangian to choose one vacuum
expectation value real, most general electroweak symmetry
violating solution can be written in a form
 \be
\langle\phi_1\rangle =\fr{1}{\sqrt{2}}\left(\begin{array}{c} 0\\
v_1\end{array}\right), \;\; \langle\phi_2\rangle
=\fr{1}{\sqrt{2}}\left(\begin{array}{c}u \\[5mm] v_2
 e^{i\xi}\end{array}\right).\label{genvac}\ee

It is useful to describe the discussed extremes  with the
aid of quantities
 \bea
 &y_1\!=\!\la\phi_1^\dagger\ra\la\phi_1\ra,\;\,
y_2\!=\!\la\phi_2^\dagger\ra\la\phi_2\ra,\;\,
y_3\!=\!\la\phi_1^\dagger\ra\la\phi_2\ra,&\nn\\
&y_3y_3^*-y_1y_2\,.&\label{yidef}
 \eea

\subsubsection{$u\neq 0$ solution, charged vacuum}

For
 \be  y_3^*y_3-y_1y_2\neq 0\;\;\mbox{we have}\;\; u\neq 0\,.
 \label{chvac}\ee
In this case the v.e.v.'s are given by equations
  \bear{c}
\lambda_1 y_1\!+\! \lambda_3y_2\!+\!\lambda_6^*
y_3^*+\lambda_6y_3=m_{11}^2/2,\\[2mm]
\lambda_2y_2\!+\! \lambda_3y_1\!+\!\lambda_7^*
y_3^*+\lambda_7 y_3=m_{22}^2/2,\\[2mm]
\lambda_4y_3^*\!+\! \lambda_5 y_3\!+\!\lambda_6 y_1+\lambda_7
y_2=m_{12}^2/2.
 \eear{chargevac}
Depending on the  parameters of potential, the extremum given by
this solution of \eqref{Eq:min-cond} describes either saddle point
or a minimum of the potential, denoted as {\it a charged vacuum},
with a heavy photon and other nonphysical properties \cite{Diaz-Cruz:1992uw}, \cite{chargebr}.

\subsubsection{$u=0$ solution, physical (neutral) vacuum}

Another solution of extremum condition \eqref{Eq:min-cond} is
realized at
 \be  y_3^*y_3-y_1y_2=0,\,\, \mbox{ which\,\, gives }\,\, u=0.\label{phvac2}\ee
The solution has a form
 \bes\label{veveq} \begin{equation}
\langle\phi_1\rangle\! =\!\dfrac{1}{\sqrt{2}}
\begin{pmatrix}
0\\[1mm] v_1
\end{pmatrix}\, \text{ and }\,
\langle\phi_2\rangle
\!=\!\dfrac{1}{\sqrt{2}}\begin{pmatrix}0 \\[1mm]
v_2 e^{i\xi}\end{pmatrix}
 \end{equation}
It satisfies a condition for $U(1)$ symmetry of
electromagnetism.

This extremum realizes minimum of potential if its parameters are
such that all eigenvalues of mass squared matrix in this extremum
point are non-negative,
sec.~\ref{secmin}. In the analysis we  consider only this very
case. (At this set of parameters the vacuum energy corresponding
to the solution \eqref{chargevac} is larger than that for the
solution \eqref{veveq} \cite{Diaz-Cruz:1992uw}, \cite{chargebr}.)

The v.e.v's $v_{1,2}$ (and therefore parameters of whole  Lagrangian)
obey SM constraint: $v_1^2+v_2^2=v^2$, with $v=(\sqrt{2}G_{\rm
F})^{-1/2}=246$~GeV. The another parameterization of these
v.e.v.'s  is also used:
\begin{equation}      \label{Eq:vevs-tanbeta}
 v_1=v\cos\beta,\quad v_2=v\sin\beta,\quad
\beta\in \left(0,\,\dfrac{\pi}{2}\right).
\end{equation}\ees

The rephasing of fields  (\ref{rephase})  shifts the phase
difference $\xi$ as
\begin{equation}
\xi\to\xi-\rho\,.\label{formxi}
\end{equation}
Therefore,  the phase difference $\xi$\  between the v.e.v.'s has
no physical sense (it was discussed e.g. in \cite{Branco}).

The  arbitrariness described by \eqref{formxi} allows to simplify
further calculations in a following way. Let us take some
Lagrangian describing our model and calculate v.e.v.'s
\eqref{veveq}. Than, by making the RPhT with $\rho=\xi$, we  get
{\em the  Lagrangian in a real vacuum form  (a
real vacuum Lagrangian)} (and {\em the  potential
in a real vacuum form}). By definition,  the relative phase of
v.e.v.'s derived from this Lagrangian equals to zero. In
accordance with eq.~(\ref{Eq:rephase}) we get now
 \bear{c}\label{newlam}
\lambda_{1-4,rv}=\lambda_{1-4},\;\,
\lambda_{5,rv}=\lambda_5e^{-2i\xi},
\\[2mm]
\lambda_{6,rv}=\lambda_{6}e^{-i\xi}, \;\,
\lambda_{7,rv}=\lambda_{7}e^{-i\xi},\\[2mm]  \vak_{rv} = \vak
e^{-i\xi}, \;\, m_{12,rv}^2= m_{12}^2e^{-i\xi},\end{array}
 \end{equation}
where we denote the  particular values of parameters of such
Lagrangian (potential) by subscript {\em rv}.

The following combinations of parameters and new quantities
are useful:
  \bear{c}
\lambda_{3,rv}+\lambda_{4,rv}+\Re\lambda_{5,rv}=\lambda_{345,rv},\\
 \dfrac{v_1}{v_2}\lambda_{6,rv}\pm
 \dfrac{v_2}{v_1}\lambda_{7,rv}=\left\{\begin{array}{c}
\lambda_{67,rv},\\
\tilde{\lambda}_{67,rv},\end{array}\right.\\[4mm]
m_{12,rv}^2=2v_1v_2(\nu+i\delta).\end{array}\label{nudeldef}
 \ee

For given $v_{1,2}$ the extremum condition (\ref{Eq:min-cond})
does not constrain $\Re m_{12,rv}^2$, while it does so for $\Im
m_{12,rv}^2$, allowing to express it via $\Im(\lambda_{5-7,rv})$:
 \be
 \delta=
\underbrace{0}_{Z_2\;sym}+\underbrace{\fr{1}{2}\,\Im\lambda_{5,rv}}_{soft}
+\underbrace{\fr{1}{2}\,\Im\lambda_{67,rv}}_{hard}\,.
\label{newlamconstr}
  \ee
Here (and in the subsequent equations) the first underbraced term
correspond to the $Z_2$ symmetric case, the second and third terms
are added to each other in the case of explicitly soft and hard
violation of $Z_2$ symmetry, respectively. In particular, in the
$Z_2$ symmetric case $ {m}_{12,rv}^2=0$ and consequently
$\Im\lambda_{5,rv}=0$.

{\it Beginning from here all expressions will be presented for the
potential in a real vacuum form, without writing explicitly the
subscript {\textit rv}}. We will explicitly comment
when other forms of Lagrangian will be discussed.\\

{\bf Remarks}

The set of real vacuum Lagrangians
forms a subspace in the entire reparametrization equivalent space
-- {\em the real vacuum Lagrangian family}.  It is pictured in
FIG.~1 by black horizontal line. In different points of this
subspace the $\tan\beta$ values are different.

\subsubsection{Our form of the potential}

It is useful for the subsequent calculations to describe the
potential in terms of  $v_1$, $v_2$ and $\nu$ instead of  three
quadratic parameters $m_{11,22}^2, m_{12}^2$ \cite{Pilaftsis:1999qt}. The eq-s
\eqref{Eq:min-cond}, \eqref{nudeldef} allow to obtain relations
 \bear{c}
m_{11}^2\!=\!\underbrace{\lambda_1v^2_1\!+\!
\lambda_{345}v_2^2}_{Z_2\; sym}\;\, \underbrace{\!-\!2\nu
v^2_2}_{soft}\;\, +\underbrace{\dfrac{v_2}{v_1}\Re\left(3v_1^2
\lambda_6\!+\!
v_2^2\lambda_7\right)}_{hard},\\[7mm]
m_{22}^2\!=\!\underbrace{\lambda_2v^2_2\!+\!
\lambda_{345}v_1^2}_{Z_2\; sym}\;\, \underbrace{\!-\!2\nu
v^2_1}_{soft}+
\underbrace{\dfrac{v_1}{v_2}\Re\left(v_1^2\lambda_6\!+\!
3v_2^2\lambda_7\right) }_{hard}.
 \end{array}\label{Eq:m11-m22}
 \ee

From these relations we obtain another form of real vacuum
potential, used in this paper:
 \bear{c}
V\!=\!
\dfrac{\lambda_1}{2}\left[(\phi_1^\dagger\phi_1)\!-\!\dfrac{v_1^2}{2}\right]^2
\!\!+\!\dfrac{\lambda_2}{2}\left[(\phi_2^\dagger\phi_2)\!
-\!\dfrac{v_2^2}{2}\right]^2\\[3mm]
+\lambda_3(\phi_1^\dagger\phi_1)(\phi_2^\dagger\phi_2)
+\lambda_4(\phi_1^\dagger\phi_2)(\phi_2^\dagger\phi_1) \\[3mm]
+\dfrac{1}{2}\left[\lambda_5(\phi_1^\dagger\phi_2)^2 +{\rm
h.c.}\right]\\[2mm] +\left\{\left[\lambda_6(\phi_1^\dagger\phi_1)
+\lambda_7(\phi_2^\dagger\phi_2)\right] (\phi_1^\dagger\phi_2)
+{\rm h.c.}\right\}  \\[2mm]
 -\dfrac{1}{2}\bigl(\lambda_{345}+2\Re
\lambda_{67}\bigr) [v_2^2(\phi_1^\dagger\phi_1) +
v_1^2(\phi_2^\dagger\phi_2)]
\\[2mm]-\,\Re[\lambda_6(\phi_1^\dagger\phi_1)
+ \lambda_7(\phi_2^\dagger\phi_2)] v_1 v_2  \\[2mm]
+ \nu(v_2\phi_1-v_1\phi_2)^\dagger(v_2\phi_1-v_1\phi_2)\\[2mm]+
2\delta\, \Im(\phi_1^\dagger\phi_2) v_1v_2
-\lambda_1\fr{v_1^4}{8}-\lambda_2\fr{v_2^4}{8}.
 \eear{GKO-pot}

In this form the quartic terms are as those in the initial
potential (\ref{Higgslagr}) but with particular values of
parameters $\lambda_i$ equal to $\lambda_{i,rv}$ (\ref{newlam}).
The mass term is determined via v.e.v.'s\ $v_1$, $v_2$ and the
parameters $\lambda_i$ plus a {\em single free} dimensionless
parameter $\nu$. The quantity $\delta \propto\Im\,m_{12}^2$ is
given by eq.~(\ref{newlamconstr}). (Sometimes instead of $\nu$ a
dimensional parameter $\mu$, defined via $\mu^2=\nu v^2$, is
used.)

In the above equation the soft $Z_2$ violating contribution  is
written as a sum of two terms, so that the variation of each of
them don't influence v.e.v.'s. The derivatives of first term
($\propto\nu$) over $\phi_k$ are equal to zero at the extremum
point $\la \phi_k\ra=v_k/ \sqrt{2}$. The second term ($\propto
\delta$) is equal to zero for real $\phi_{1,2}$, independently on
their absolute values. This decomposition is less transparent in
Lagrangians with $\xi\neq 0$.

The vacuum energy density given by minimum \eqref{veveq} is equal
to
 \bear{c}
E_{vac}\equiv V\left(\la\phi_1\ra,\la\phi_2\ra\right)
=-\fr{\lambda_1v_1^4}{8}-\fr{\lambda_2v_2^4}{8}\\-
\lambda_{345}\fr{v_1^2v_2^2}{4}
-\Re(\lambda_6v_1^2+\lambda_7v_2^2)\fr{v_1v_2}{2}\,.
 \eear{nvacE}

{\bf Remark}

The transformation \eqref{reparam} with $\tau=0$, $\theta=-\beta$
or $\theta=\pi/2-\beta$ gives ($v_1=v$, $v_2=0$) or  ($v_1=0$,
$v_2=v$), respectively. The sets of the obtained Lagrangians form
{\em a Higgs basis Lagrangian families}, they are pictured as grey
vertical strips in the reparameterization equivalent Lagrangian
space presented in FIG.~1.  These cases  cannot be described by
our potential \eqref{GKO-pot} since some of the coefficients
\eqref{nudeldef}, \eqref{newlamconstr}, used at the transformation
to this form, are singular at $v_2\to 0$ or $v_1\to 0$. For both
these cases $\sin2\beta=0$. Therefore, our analysis based on the
potential \eqref{GKO-pot} is valid only for Lagrangians with
 \be
 \sin2\beta\neq 0\,\label{Higgsbexcl}
 \ee
(domain of entire reparametrization equivalent space of FIG.~1
between two grey strips). Some results for the Higgs basis
Lagrangian can be found in \cite{higgsbasis,Branco,Higgsrep}.

\section{Physical Higgs sector}\label{secphysHig}


The fields $\phi_{1,2}$ change under the transformation \eqref{reparam}.
We introduce
now the,  in principle, observable Higgs fields and their couplings.
These fields and couplings are evidently reparametrization
independent. (The reparametrization dependent are parameters
describing the transformation  to this physical basis, see below.)

A standard decomposition of the fields $\phi_{1,2}$ in terms of
physical fields is made via
\begin{equation}
\phi_1\!=\!\begin{pmatrix} \varphi_1^+ \\
\dfrac{v_1+\eta_1+i\chi_1}{\sqrt{2}}
\end{pmatrix}, \;\,
\phi_2=\begin{pmatrix} \varphi_2^+ \\
\dfrac{v_2+\eta_2+i\chi_2}{\sqrt{2}}
\end{pmatrix}.\label{videf}
\end{equation}
At $\vak=0$ such decomposition leads to a diagonal form of kinetic
terms for new fields $\varphi_i^+,\,\chi_i,\,\eta_i$, while the
corresponding mass matrix is off-diagonal. The mass-squared matrix
can be transformed to the block diagonal form by a separation of
the massless Goldstone boson fields,
$G^0=\cos\beta\,\chi_1+\sin\beta\,\chi_2$ and
$G^\pm=\cos\beta\,\varphi_1^\pm+\sin\beta\,\varphi_2^\pm$, and the
charged Higgs boson fields $H^\pm$:
 \be
H^\pm=-\sin\beta\,\varphi_1^\pm+\cos\beta\,\varphi_2^\pm\, ,
\label{physfieldspm}
\end{equation}
with  the mass squared  equal to
\begin{equation}   \label{Eq:mch}
M_{H^\pm}^2= \left[\nu-\dfrac{1}{2}( \lambda_4+\Re \lambda_5+\Re
\lambda_{67})\right]v^2.
\end{equation}

\subsection{Neutral Higgs sector. General introduction}

By definition $\eta_{1,2}$ are the standard $C$-- and $P$-- even (scalar)
fields. The  field
\begin{equation}
A=-\sin\beta\,\chi_1+\cos\beta\,\chi_2\,, \label{physfields}
 \end{equation}
is $C$--odd (which in the interactions with fermions behaves as a
$P$-- odd particle, i.e. a pseudoscalar). In other words, the
$\eta_{1,2}$ and $A$ are fields with opposite CP parities (see
e.g. \cite{Hunter} for details). (Note that sometimes  the set
$\eta_1$, $\eta_2$ and $A$ is called  the weak basis
\cite{Branco}.)

The decomposition (\ref{videf}) results in the (symmetric)
mass--squared matrix $\cal M$ in the $\eta_1$, $\eta_2$, $A$ basis
 \bes\label{Eq:M3by3}\begin{equation}
{\cal M}=\begin{pmatrix}
M_{11} & M_{12} & M_{13} \\
M_{12} & M_{22} & M_{23} \\
M_{13} & M_{23} & M_{33}
\end{pmatrix},
\end{equation}
 with
\begin{eqnarray}
 &M_{11}=\left[c_\beta^2\,\lambda_1 +s_\beta^2\,\nu+
s_\beta^2
\Re(\dfrac{\lambda_{67}}{2}+\tilde{\lambda}_{67})\right]v^2,&
\nonumber \\
&M_{22}=\left[s_\beta^2\,\lambda_2+c_\beta^2\,\nu +c_\beta^2
\Re(\dfrac{\lambda_{67}}{2}-\tilde{\lambda}_{67})\right]v^2 ,&
\nonumber \\
&M_{33}=\left[\nu-\Re(\lambda_5-\dfrac{1}{2}\lambda_{67})
\right]v^2,&
\label{Eq:M_ij} \\
&M_{12}=-\left[\nu-\lambda_{345}
-\dfrac{3}{2}\Re\lambda_{67}\right]c_\beta s_\beta v^2,
&\nonumber\\
&M_{13}=-\left[\delta
+\fr{1}{2}\Im\tilde{\lambda}_{67}\right]s_\beta
v^2,&\nn\\
& M_{23}=-\left[\delta
-\fr{1}{2}\Im\tilde{\lambda}_{67}\right]c_\beta v^2,&\nonumber
\end{eqnarray}
where we use abbreviations $c_\beta=\cos\beta$,
$s_\beta=\sin\beta$. As we discuss below $M_{33}$ is equal to the
mass squared of the CP--odd Higgs boson  in the CP conserving
case, namely
 \begin{equation}
M_A^2=M_{33}= \left[\nu-\Re(\lambda_5-\dfrac{1}{2} \lambda_{67})
\right]v^2 \label{Eq:MA}.
 \end{equation}\ees

The masses squared $M_i^2$ of the physical neutral states $h_{1-3}$
are eigenvalues of the matrix $\cal M$. These states are obtained
from fields $\eta_1,\,\eta_2,\,A$ by a unitary transformation $R$
which diagonalizes the  matrix $\cal M$:
 \bear{c} 
\begin{pmatrix}
h_1 \\ h_2 \\ h_3
\end{pmatrix}
= R
\begin{pmatrix}
\eta_1 \\ \eta_2 \\ A
\end{pmatrix}\,,\\[6mm] \mbox{ with }\;\; R{\cal
M}R^T=diag(M_1^2,\,M_2^2,\,M_3^2)\,.
 \end{array}\label{Eq:R33}
 \ee
The diagonalizing matrix $R$ can be  written as a product of three
rotation matrices described by three Euler angles $\alpha_i\in
(0,\pi)$ (we define $c_i=\cos\alpha_i$, $s_i=\sin\alpha_i$):
    \bes\label{Eq:R-angles}
  \bear{c}
R=R_3R_2R_1,\quad R_1=\begin{pmatrix}
c_1 & s_1 & 0 \\
-s_1 & c_1 & 0 \\
0         &       0         & 1
\end{pmatrix},\\[7mm]
R_2\! =\!\begin{pmatrix}c_2&0&s_2\\
0&1&0\\-s_2&0&c_2\end{pmatrix},\; R_3\!=\!\begin{pmatrix}1&0&0\\
0&c_3&s_3\\ 0&-s_3&c_3\end{pmatrix},
 \eear{defangles}
\bear{c}
     R=\begin{pmatrix}
R_{11} & R_{12} & R_{13} \\
R_{21} & R_{22} & R_{23} \\
R_{31} & R_{32} & R_{33}
\end{pmatrix}\equiv\\[7mm]
\begin{pmatrix}
c_1\,c_2 & c_2\,s_1 & s_2 \\
\!-\! c_1\,s_2\,s_3\! -\! c_3\,s_1
& c_1\,c_3\! -\! s_1\,s_2\,s_3 & c_2\,s_3 \\
\!-\! c_1\,c_3\,s_2\! +\! s_1\,s_3 &\! -\! c_1\,s_3\! -\!
c_3\,s_1\,s_2 & c_2\,c_3
\end{pmatrix}\,.
\eear{}\\\ees We adopt the convention for  masses that
$M_2\ge M_1$, but shall {\it not} require any other
ordering.

In general, the obtained Higgs eigenstates $h_i$ (\ref{Eq:R33})
have no definite CP parity  since they are mixtures of fields
$\eta_{1,2}$ and $A$ having the opposite CP parities. This
provides a CP nonconservation within the Higgs sector. The
interaction of these Higgs bosons with matter  explicitly violates
the CP--symmetry. Such  mixing  (and  violation  of CP) is absent
if
$M_{13}=M_{23}=0$.\\

{\bf Remarks}

\bu  Note that in the case where there is an exact $Z_2$ symmetry
and  $\lambda_5=0$ there appears an additional Peccei--Quinn
symmetry. Then $A$ is a massless Goldsone-like boson, $M_A=0$. The
spontaneous violation of this symmetry results in a light particle
with mass, which is generated due to non-perturbative effects.

\bu For the basic Higgs Lagrangian (\ref{Higgslagr}) in the case
of a soft violation of $Z_2$ symmetry and Model II or I for the
Yukawa interaction (see below),  the perturbative corrections give
no counter terms violating the $Z_2$ symmetry in a hard way.
Therefore, with a suitable renormalization procedure, the  mixed
kinetic terms don't appear  in the Lagrangian in $h_{1-3}$ basis
(the rotation (\ref{Eq:R33}) keeps kinetic term diagonal in all
orders). At the same time, the  mass terms and mixing angles
$\alpha_i$ change due to the renormalization. Some aspects of this
procedure were discussed in \cite{Pilaf97}.

\subsection{Diagonalization of the scalar CP-even sector}\label{secscalar}

It is useful to start with the diagonalization of scalar $\la
12\ra$\, sector of matrix $\cal M$ which is given by the rotation
matrix $R_1$. It results in the neutral, CP-even Higgs fields
which we denote as $h$ and $(-H)$, while the CP--odd field $A$
remains unmixed. (Sign minus at $H$ is needed in order to match a
standard convention used for CP-conserving case, see e.g.
\cite{Hunter}.) We got
 \bear{c}  \label{Eq:diagonal-2times2}
\begin{pmatrix}
h \\ -H \\ A
\end{pmatrix}= R_1\begin{pmatrix}
\eta_1 \\ \eta_2 \\ A
\end{pmatrix}
 \; \mbox{ with }\\[6mm]
 R_1{\cal M}R_1^T={\cal M}_1
 \equiv
\begin{pmatrix}
M_h^2 &   0   & M_{13}' \\
  0   & M_H^2 & M_{23}' \\
M_{13}' & M_{23}'   & M_A^2
\end{pmatrix}.\end{array}
 \end{equation}
with $ M_{13}',M_{23}'$ given in eq. ~\eqref{mixCPgen}.

{\it Let us stress that in the general CP nonconserving case the
states $h$, $H$ and $A$ have no direct physical sense, they are
only  subsidiary concepts  useful in the calculations and
discussions.} In the case of CP conservation (realized for
$M_{13}'=M_{23}'=0$) the fields $h$, $H$ and $A$ represent
physical Higgs bosons: $h_1=h$, $h_2=-H$, $h_3=A$. This   is why
we  use instead of $\alpha_1$ the mixing angle $\alpha\in
(-\pi/2,\,\pi/2)$,
 \be
 \alpha=\alpha_1-\pi/2\,,
 \ee
which is customary for the CP--conserving case. Using this angle
we get
 \bear{c}   H=
\cos\alpha\,\eta_1+\sin\alpha\,\eta_2,\\[1mm]
h=-\sin\alpha\,\eta_1+\cos\alpha\,\eta_2 .\end{array}
\label{Eq:define-alpha}
 \end{equation}
The diagonalization of the respective $\la 12\ra$ corner of
mass-squared matrix $\cal M$ (\ref{Eq:M3by3}) results in
 \bear{c}
M_{h,H}^2=\dfrac{1}{2}\left(M_{11} +M_{22}\mp {\cal
N}\right),\\[4mm]
{\cal N}=\sqrt{(M_{11}-M_{22})^2+4M_{12}^2}.
\end{array}\label{massevendiag}
\end{equation}

The following expressions for angles are  useful in some
applications:
 \bear{c}
\cos2\alpha=\dfrac{M_{11}-M_{22}}{M_H^2-M_h^2} ,\\[3mm]
\sin2\alpha=\dfrac{2M_{12}}{M_H^2-M_h^2}\, , \\[3mm]
\dfrac{\sin2\alpha}{\sin 2\beta}=\dfrac{(\lambda_{345}-\nu+3/2\Re
\lambda_{67})v^2} {M_H^2-M_h^2}.\end{array} \label{Eq:mHh}
 \end{equation}

\subsection{Complete diagonalization}\label{refcompl}

The above diagonalization keeps, in general, two off-diagonal
elements in matrix ${\cal{M}}_1$
(\ref{Eq:diagonal-2times2}):
 \bear{c}
M_{13}'=c_1M_{13}+s_1M_{23}=\\[2mm]-\left[\delta
\cos(\beta+\alpha)-\fr{Im\tilde{\lambda}_{67}}{2}
\cos(\beta-\alpha) \right]v^2,\\[2mm]
\quad M_{23}'=-s_1M_{13}+c_1M_{23}=\\[2mm]\left[\delta
\sin(\beta+\alpha)+\fr{Im\tilde{\lambda}_{67}}{2}
 \sin(\beta-\alpha)\right]v^2.\end{array}
\label{mixCPgen}
\end{equation}
If  at least one of these off diagonal terms differs from zero,
the additional diagonalization is necessary, and the mass
eigenstates, being admixtures of  CP--even and CP--odd states,
violate the CP symmetry. In this case we express the
physical Higgs boson states $h_{1-3}$ via $h$, $H$, $A$:
 \bea
&\begin{pmatrix} h_1 \\h_2 \\ h_3
\end{pmatrix}
= R_3R_2\begin{pmatrix} h \\ -H \\ A
\end{pmatrix}
\;\mbox{ with}& \label{semidiag}
\\[3mm]
&R\,{\cal M}\,R^T= R_3R_2{\cal M}_1R_2^TR_3^T=
\begin{pmatrix}
M_{1}^2 &    0     &   0\\
0        & M_{2}^2 & 0 \\
0        &  0       & M_{3}^2
\end{pmatrix}.&\nn
 \eea

The squared masses $M_i^2$ in Eq.(\ref{semidiag}) are the
eigenvalues of the mass-squared matrix $\cal M$ (\ref{Eq:M3by3}),
i.e. they are roots of the corresponding cubic equation (see
solution, e.g., in ref.~\cite{Dubinin:2002nx}). Note, that the
trace of mass-squared matrix does not changed under the unitary
transformations. Therefore, we have {\it mass sum rule}
 \bear{c}
M_1^2+M_2^2+M_3^2=M_h^2+M_H^2+M_A^2\\[1mm]=M_{11}+M_{22}+M_{33}\,.
\end{array}\label{massSR}
\end{equation}

The relation \eqref{semidiag} allows to discuss the general CP
violating case in terms customary for the CP conserving one, i.e.
with {\it parameters} $M_H,\,M_h$, $M_A$ and $\alpha$. The angles
$\alpha_2$, $\alpha_3$ describe mixing of the CP--even states
($h$, $H$) with the CP--odd state $A$.

\subsection{\bf Condition for CP violation.}

The eq.~\eqref{Eq:diagonal-2times2} shows explicitly that the CP
does not violated if and only if simultaneously
 \be
 M'_{13}=0\,,\quad  M'_{23}=0\,.\label{CPconscond}
 \ee
If these conditions are fulfilled, one can express the
ratio $\Im\tilde{\lambda}_{67}/\delta$ in terms of angles
$\beta$ and $\alpha$ by two relations which contradict each
other at $\sin2\beta\neq 0$ \eqref{Higgsbexcl}. Therefore,
the CP violation in Higgs sector is absent if and only if
$\Im\tilde{\lambda}_{67}=0$ and $\delta\propto
\Im(m_{12}^2)=0$. From the \eqref{newlamconstr} it follows
that the CP violation is absent if all coefficients in
potential of a real vacuum form are real. Simple but
cumbersome  calculation shows that similar conclusion is
valid also for the potential in a Higgs basis form, i.e.
for $\sin2\beta=0$. In other words, CP symmetry in Higgs
sector is not violated if among different reparametrization
equivalent potentials a potential with all real
$\lambda_i$, $m^2_{ij}$ parameters can be found.

Vice versa, the complexity of some parameters of the potential in
{\it a real vacuum form} is a sufficient condition for CP violation in
the Higgs sector. For an arbitrary form of Lagrangian (in entire
reparametrization space) the necessary and sufficient condition
for CP violation in the Higgs sector can be written as complexity
of some of combinations (which are invariant under RPhT, see
\eqref{Eq:rephase})
 \be
\lambda_5^*(m_{12}^2)^2\,,\quad \lambda_6^*m_{12}^2\,,\quad
\lambda_7^*m_{12}^2\,.
 \label{newcond}
 \ee
Each this quantity is not reparametrization invariant one
but they are very simple. (The condition for the case of
explicit soft $Z_2$ violated potential was obtained in
\cite{Froggatt:1992wt}). More complex form of
conditions for CP violation can be written via invariants
of RPaT \cite{invappr}, \cite{GH05}.

\subsection{Various cases  of CP violation}\label{secsomecase}

Here we present various cases of  CP violation.

$\bullet \,\,\,$ If {\bf\bm$\delta=0$ and $\Im
\tilde{\lambda}_{67}=0$}, {CP symmetry  is  not violated}. The $h$,
$H$ and $A$ are physical Higgs bosons, with  masses  given by
eqs.~(\ref{massevendiag}) and (\ref{Eq:MA}), and
$\alpha_2=\alpha_3=0$.

$\bullet$ If {\bf\bm  $\vep_{13}\equiv|M_{13}'/ (M_A^2-M_h^2)|\ll
1 $} the  Higgs boson $h_1$ practically coincides with $h$
($\alpha_2\approx 0$). The interaction of $h_1$ with other
particles respects CP--symmetry (with an accuracy
$\sim\vep_{13}$). The diagonalization of the residual $\la 23\ra$
corner of mass-squared  matrix \eqref{Eq:diagonal-2times2} with
the aid of rotation matrix $R_3$ (\ref{defangles}) gives states
$h_2$ and $h_3$. They are  superpositions of $H$ and $A$ states
with potentially large mixing angle $\alpha_3$:
 \bes\label{tanalph}\be
\tan2\alpha_3\approx \dfrac{-2M'_{23}}{M_A^2-M_H^2}\,,\;\;
\alpha_2\approx 0.\label{tanalph3}
 \ee
If $M_A\approx M_H$, the CP violating mixing can be strong even at
small but nonzero $|M'_{23}|/v^2$. The states $h_2$ and $h_3$ have
no definite CP parity and the mass difference $|M_2^2-M_3^2|$ is
larger than $|M_H^2-M_A^2|$.\fn{In this case one can hope to
separate $h_2$ and $h_3$ at muon collider (see \cite{Godang} for
the case without mixing) and to measure mixing angle $\alpha_3$
via measuring of difference in effects of CP violation in the
corresponding two peaks.}  For example, at $M_H\approx 300$ GeV,
$|M_H-M_A|\approx 5$ GeV and $M'_{23}\approx 0.02\,v^2$ we have
$|M_2^2-M_3^3|\approx 25$ GeV, $\sin2\alpha_3\approx 0.8$.

At the growth of $M_H\approx M_A$ the proper widths of $H$ and $A$
become large so that the $H$ and $A$ peaks overlap strong. In such
case the tree approximation may be too rough for a reliable
calculation of masses $M_i$ and mixing angles. Therefore one
should supplement the mass-squared matrix (\ref{Eq:M3by3}) by a
(complex) matrix of Higgs polarization operators as it is
customary in the description of low energy phenomena (see
discussion in sec.~\ref{secdecoupl}).

$\bullet$  If {\bf\bm  $ \vep_{23} \equiv
|M_{23}'/(M_A^2-M_H^2)|\ll 1$}, the Higgs boson $h_2$ practically
coincides with $-H$ ($\alpha_3\approx 0$). The interaction of
matter with $h_2$ does not violate CP--symmetry. Similarly to the
previous case, the diagonalization of  the $\la 13 \ra$ part of
mass-squared matrix \eqref{Eq:diagonal-2times2}, with the aid of
rotation matrix $R_2$ (\ref{defangles}), gives states $h_1$ and
$h_3$. They are  superpositions of $h$ and $A$ states with
potentially large mixing angle $\alpha_2$:
 \be
 \tan2\alpha_2\approx
\dfrac{-2M'_{13}}{M_A^2-M_h^2}\,,\;\; \alpha_3\approx 0.
\label{tanalph2}
 \ee \ees
Similarly to the previous case, if $M_A\approx M_h$, the CP
violating mixing can be strong even at small $M'_{13}/v^2$.

$\bullet$ {\bf Case of a weak CP violation} combines both described
above cases \eqref{tanalph}. If both $|M'_{13}|\ll|M_A^2-M_h^2|$
and $|M'_{23}|\ll|M_A^2-M_H^2|$ the CP--even states $h$, $H$ are
weakly mixed with the CP--odd state $A$, and parameters $\alpha_2$
and $\alpha_3$ are simultaneously small:
 \begin{subequations}\label{weakCPgen}
 \bear{c}\label{weakCPsi}
\tan\alpha_2\approx s_2\approx\dfrac{ -M_{13}'}
{M_A^2-M_h^2}\approx \alpha_2\,,\\[3mm] \tan\alpha_3\approx
s_3\approx\dfrac{ -M_{23}'} {M_A^2-M_H^2}\approx \alpha_3\\[3mm]
(|\alpha_2|,|\alpha_3|\ll 1).\end{array}
 \ee

To the second order in $s_2$ and $s_3$ the corresponding masses
are
 \bear{c}
M_1^2=M_h^2-s_2^2(M_A^2-M_h^2),\\[1mm] M_2^2=M_H^2-s_3^2(M_A^2-M_H^2),
 \end{array}\label{weakCP1}\ee
with $M_3$ given by the sum rule (\ref{massSR})\,.

In the particular case of {\it a soft violation of $Z_2$ symmetry}
we have
 \be
 s_2\!\approx\! \delta \,
\dfrac{\cos(\beta+\alpha)}{M_A^2\!-\!M_h^2}v^2,\; s_3\!\approx\!
-\delta\,
\dfrac{\sin(\beta+\alpha)}{M_A^2\!-\!M_H^2}v^2.\label{weakCPssi}
 \ee\ees

\bu The case of {\bf intense coupling regime} with $M_A\approx
M_h\approx M_H$ \cite{intense} may also give strong CP violating
mixing even with small both $\delta$ and $\Im \tilde{\lambda}_{67}$.\\

{\bf Remark}

Note that in MSSM, etc. CP symmetry can be violated by interaction
of Higgs fields with different scalar squarks, etc. In this case
the mixed polarization operators $\Im \Pi_ {HA}$ and $\Im\Pi_
{hA}$ appear leading to the CP violation in Higgs sector even for
the CP conserving Higgs potential. This violation can be visible
if $H$ and $A$ or (and) $h$ and $A$ are almost degenerate (see
e.g. \cite{EllPilaf} and references therein).

\subsection{Couplings to gauge bosons}

The gauge bosons $V$ ($W$ and $Z$) couple only to the CP--even
fields $\eta_1$, $\eta_2$. For the physical Higgs bosons $h_i$
(\ref{Eq:R33}) one obtains simple expressions for their couplings,
which in terms of the relative couplings (\ref{Eq:gj}) read
 \bes\label{Eq:W}
 \bear{c}
\chi_V^{(i)}\!=\!\cos\beta\, R_{i1}\! +\!\sin\beta\, R_{i2}, \;\,
 V=W\mbox{ or } Z. \end{array}
\label{Eq:WW}
\end{equation}

Note that due to  unitarity of the transformation matrix $R$, the
following sum rule takes place \cite{sumrule}:
 \be
 \sum\limits_{i=1}^{3} (\chi_V^{(i)})^2=1\,.\label{srW}
 \ee\ees

In particular, in the case of weak violation of the CP symmetry
considered above, with $s_2$, $s_3$ given by
eqs.~(\ref{weakCPsi}), we obtain
 \bear{c}\label{Eq:WCPweak}
 \chi_V^{(1)}=\sin(\beta-\alpha),\;\,
 \chi_V^{(2)}=-\cos(\beta-\alpha),\\
 \chi_V^{(3)}=-s_2\sin(\beta-\alpha)+s_3\cos(\beta-\alpha).\end{array}
 \ee

\subsection{Higgs self-couplings }\label{secself}

The decomposition of the scalar fields $\phi_{1,2}$ in terms of
physical fields $h_i$ allows to identify the trilinear and quartic
couplings among them via parameters of Lagrangian and elements of
mixing matrix (\ref{Eq:R-angles}). They were obtained in
\cite{Choi:1999uk,Carena:2002bb,Akhmetzyanova:2004cy}. For
completeness, we present them for our specific form of Lagrangian
(\ref{GKO-pot}) in the Appendix.

In the case of  soft $Z_2$ symmetry violation in the CP
conservation case  these equations  simplify and we present in the
Appendix self--couplings in this particular case as well.

For this very case we present  two  useful forms
for triliniear couplings. First, we express these couplings in
terms of masses and mixing angles $\alpha$ and $\beta$.  Second,
for the case of Model II for Yukawa interaction (see below) we
find expressions for these trilinear couplings in terms of masses
and relative couplings to gauge bosons and quarks (and the
parameter $\nu$).

\section{Yukawa interactions}\label{secYuk}


\subsection{General discussion}\label{secYUkgen}

In the general case the Yukawa Lagrangian reads \cite{Branco}
  \bear{c}
-{\cal L}_{\rm Y} =\bar Q_L [(\Gamma_1\phi_1+\Gamma_2\phi_2)
d_R\\[2mm]
+(\Delta_1\tilde\phi_1+\Delta_2\tilde\phi_2) u_R] +{\rm h.c.},
 \end{array}\label{Eq:Yuk00}
 \ee
with  similar terms for the  leptons. Here, $Q_L$ refers to
the 3-family vector of the left-handed quark doublets,
whereas $d_R$ and $u_R$ refer to  the 3-family vectors of
the the right-handed field singlets (with $q_{\rm L}=
(1-\gamma_5)q/2$ and $\tilde\phi_a=i\tau_2\phi_a^{\dagger
{\rm T}}$). The Yukawa matrices $\Gamma$ and $\Delta$ are
3--dimensional matrices in the family space with generally
complex elements (Yukawa parameters).

Obviously the transformation (\ref{reparam}) induces changes in the elements
of matrices $\Gamma_i$ and $\Delta_i$. In particular, {\it the
rephasing invariance} is extended to the full Higgs + Yukawa Lagrangian space
if
one supplements the transformations (\ref{rephase}) of fields
$\phi_{1,2}$ by the following transformations of fermion fields
 \begin{subequations}\label{rephasingYukgen} \bear{c}
 Q_{Lk}\to Q_{Lk}e^{i\tau_{qk}},\\[2mm]
 d_{Rk}\to d_{Rk}e^{i(\tau_{qk}+\tau_{dk})},\\[2mm]
 u_{Rk}\to u_{Rk}e^{i(\tau_{qk}+\tau_{uk})}.\end{array}\label{rephasingYukf}
  \ee

The corresponding transformations of the parameters of Yukawa
Lagrangian supplementing the transformations (\ref{Eq:rephase}) are
 \bear{c}
 \Gamma_1\,\to\,\Gamma_1 \begin{pmatrix}
 e^{i\tau_{d1}}\\ e^{i\tau_{d2}}\\ e^{i\tau_{d3}}\end{pmatrix}e^{-i\rho_1},
 \;
 \Delta_1\,\to\,\Delta_1 \begin{pmatrix}
 e^{i\tau_{u1}}\\ e^{i\tau_{u2}}\\
 e^{i\tau_{u3}}\end{pmatrix}e^{i\rho_1},\\[5mm]
 \Gamma_2\,\to\,\Gamma_2 \begin{pmatrix}
 e^{i\tau_{d1}}\\ e^{i\tau_{d2}}\\
 e^{i\tau_{d3}}\end{pmatrix}e^{-i\rho_2},
\; \Delta_2\,\to\,\Delta_2 \begin{pmatrix}
 e^{i\tau_{u1}}\\ e^{i\tau_{u2}}\\
 e^{i\tau_{u3}}\end{pmatrix}e^{i\rho_2}. \end{array}\label{rephasingYuk}
 \ee  \ees

An existence  of the off-diagonal (in family index) terms
in the Yukawa  matrices results in the flavor-changing
neutral-currents (FCNC). The rephasing invariance under the
transformations (\ref{rephase}), (\ref{rephasingYuk})
allows to make real the diagonal elements of only one
matrix $\Gamma$ and one matrix $\Delta$. Complex values of
the other elements of matrices $\Gamma_{1,2}$ and
$\Delta_{1,2}$ can result in the complex values of
one--loop corrections to some $\lambda$'s and in
consequence to the CP violation in the Higgs sector
discussed above (even for real bare coefficients $m_{12}^2$
and $\lambda$'s). In the latter case, the corresponding CP
violating terms should be included in the Higgs Lagrangian
in order to have a standard multiplicative
renormalizability.

Note, that in the case when simultaneously $\Gamma_1\neq 0$
and $\Gamma_2\neq 0$ or $\Delta_1\neq 0$ and $\Delta_2\neq
0$ (i.e. right-handed fermion of the type $d_R$ or $u_R$
interacts with both fields $\phi_1$ and $\phi_2$), the
counter terms corresponding to the  one-loop corrections to
the Higgs Lagrangian contain operators of dimension 4,
which violate $Z_2$ symmetry (\ref{Eq:Z2-symmetry}) in a
hard way. They contribute to the renormalization of
parameters $\vak$, $\lambda_6$ and $\lambda_7$
\cite{Ginzburg1977}, \cite{Froggatt:1992wt}. Therefore, to
have only the soft violation of $Z_2$ symmetry (to prevent
$\phi_1\leftrightarrow\phi_2$ transitions at small
distances), one demands that
\cite{Glashow:1977nt,Paschos:1976ay}
 \bear{c}
\mbox{\it each right-handed fermion couples to}\\ \mbox{\it  only
one scalar field, either $\phi_1$ or $\phi_2$.}
 \eear{Yukprop}

The case $\Gamma_2=\Delta_1=0$ with diagonal $\Gamma_1$,
$\Delta_2$ corresponds to the well known Model II, while
$\Gamma_2=\Delta_2=0$ -- to the Model I (see\, e.g.
\cite{Hunter}). For the Lagrangian having simultaneously
Model II and soft $Z_2$ violated form  these properties
(i.e. soft $Z_2$ violation and \eqref{Yukprop}) are stable
under the radiative corrections. Note that general RPaT
makes these properties of Lagrangian hidden.

If for a given physical system both the Model II (or Model I) and
the soft $Z_2$ violating Lagrangians  exist but don't coincide,
the radiative corrections transform a
Lagrangian to that with a true hard violation of $Z_2$ symmetry.
In this case only a general model like Model III and with true
hard violation of the $Z_2$ symmetry is renormalizable. We don't
study this case considering it as {\it unnatural}.

\subsection{Model II}\label{secmodII}\vspace{-5mm}

We limit ourselves to the case when the physical reality
allows for the description of Higgs--fermion interaction in
a form, where the fundamental scalar field $\phi_1$ couples
to $d$-type quarks and charged leptons $\ell$, while
$\phi_2$ couples to $u$-type quarks (we take neutrinos to
be massless) -- the Model II Lagrangians, which are
represented by a crossed vertical strip in FIG.~1.

Using matrices $\Gamma_1=diag(g_{d1},g_{d2},g_{d3})$
and\linebreak[4] $\Delta_2 =diag(g_{u1},g_{u2},g_{u3})$, we get
 \bear{c}
  -{\cal L}_Y^{II}= \sum_{k=1,2,3}g_{dk}\bar{Q}_{Lk}
\phi_1 d_{Rk}\\[2mm] + \sum_{k=1,2,3}g_{uk} \bar{Q}_{Lk} \tilde\phi_2
u_{Rk}\\[2mm]+ \sum_{k=1,2,3}g_{\ell k} \bar{\ell}_{Lk} \phi_1
\ell_{Rk} +{\rm h.c.}
 \end{array}
\label{YukII}
\end{equation}

Certainly, the general RPaT \eqref{newlagr} transforms Yukawa
Lagrangian to a form where this basic definition of Model II
cannot be seen. That are  {\it hidden Model II forms of
Lagrangian}. In entire reparametrization equivalent space these
Lagrangians form a family shown by a crossed vertical strip in
FIG.~1 (in this figure we suggest that this family coincides with
soft $Z_2$ symmetry violated family).

The suitable choice of phases in transformations
(\ref{Eq:rephase}) and \eqref{rephasingYuk} eliminates phase
difference of vacuum expectation values and makes all Yukawa
parameters real. It gives {\em a Model II Lagrangian in a real
vacuum  form}. We will use such Lagrangian below. It corresponds
to the intersection of crossed and black strips in FIG.~1.

As it was written above, different forms of Lagrangian can have
different values of $\tan\beta$.   To underline that we use the
mentioned Lagrangian, we will supply (only in this section)
quantity $\beta$ in this case by a subscript II, $\beta\to
\beta_{II}$.

Since v.e.v.'s of scalar  fields are responsible for the fermion
mass similarly as in the SM, the relative Yukawa couplings of the
physical neutral Higgs bosons $h_i$  (\ref{Eq:gj})  are identical
for all $u$--type and for all $d$--type quarks (and charged
leptons). They can be expressed via elements of the rotation
matrix $R$ (\ref{Eq:R33}):
 \bear{c}\label{Eq:chi-ud}
\chi_u^{(i)}= \dfrac{1}{\sin\beta_{II}}[R_{i2}-i\cos\beta_{II}\,R_{i3}],\\
\chi^{(i)}_d
=\dfrac{1}{\cos\beta_{II}}[R_{i1}-i\sin\beta_{II}\,R_{i3}].
 \end{array}\ee
(Note that e.g. the interaction $\bar{d}_L(g_1+ig_2)d_R+h.c$ reads
for the Dirac fermions as $\bar{d}(g_1-i\gamma_5g_2)d$.)

In the particular case of weak CP violation (with small
$s_2$, $s_3$ (\ref{weakCPgen})) these relative couplings,
together with the corresponding ones to gauge bosons, are
presented in Table~\ref{couptab}.

For the interaction of the charged Higgs bosons e.g. with
$t$-quark, the Lagrangian (\ref{YukII}) gives
 \bear{c}
{\cal L}_{H^- tb}=\fr{M_t}{v\sqrt{2}}\cot\beta_{II}\;
\bar{b}(1+\gamma^5)H^-t\\[2mm] + \fr{M_b}{v\sqrt{2}}\tan\beta_{II}\;
\bar{b}(1-\gamma^5)H^-t + h.c.\end{array}\label{Hcharged}
 \ee
  \begin{widetext}\begin{center}\begin{table}[hbt]
\caption{\it  Basic relative couplings in the weak
CP-conserving 2HDM (II). In the upper lines results for the
case with no CP violation and in lower lines the
corresponding corrections $\propto s_2,\,s_3$ are
presented. }
\begin{center}
\begin{tabular}{|l|c|c|c|}
\hline &$\chi_V$&$\chi_u$&$\chi_d$ \\\hline &&&\\
 $h_1$&  \bra{c}\sin(\beta_{II}-\alpha)\\( +0)\era  &
\bra{c}\fr{\cos\alpha}{\sin\beta_{II}}\\ ( -is_2 \cot \beta_{II})
\era &
 \bra{c}-\fr{\sin\alpha}{\cos\beta_{II}}\\
( -is_2 \tan \beta_{II}) \era\\\hline &&&\\

$h_2$&\bra{c}-\cos(\beta_{II}-\alpha)\\ (+0)\era
&\bra{c}-\,\fr{\sin\alpha}{\sin\beta_{II}}\\(-is_3 \cot
\beta_{II}) \era&
\bra{c}-\,\fr{\cos\alpha}{\cos\beta_{II}}\\(-is_3 \tan \beta_{II})
\era\\\hline

$h_3$&\bra{c}0\\(-s_2\sin{(\beta_{II}-\alpha)}+s_3\cos{(\beta_{II}-\alpha)})\era&
\bra{c}-i\cot\beta_{II}\\
 ( - s_2\dfrac{\sin
\alpha}{\sin \beta_{II}} +s_3\dfrac{\cos \alpha}{\sin\beta_{II}})
\era & \bra{c}-i\tan\beta_{II}\\(+
s_2\dfrac{\sin\alpha}{\cos\beta_{II}}  +
s_3\dfrac{\cos\alpha}{\cos\beta_{II}})\era \\ \hline
\end{tabular}
\end{center}

\label{couptab}
\end{table}\end{center}\end{widetext}

It is useful to express the relative coupling of the
neutral scalar $h_i$  to the charged Higgs boson, in the
cases of weak CP violating  and soft $Z_2$-violation, via
the relative couplings of this neutral Higgs boson to the
gauge bosons and fermions:
 \bear{c}
\chi_{H^\pm}^{(i)}
=\left(1-\fr{M_i^2}{2M_{H^\pm}^2}\right)\chi_V^{(i)}\\[3mm]
+\fr{M_i^2-\nu v^2}{2M_{H^\pm}^2}
\Re(\chi_u^{(i)}+\chi_d^{(i)}).
 \end{array}\label{b2d3}
\end{equation}

\subsection{ Set of useful relations  in Model~II}\label{secpat}

The  unitarity of the mixing matrix $R$ allows to obtain a number
of relations \cite{sumrule,Ginzburg:2001ss,Ginzburg:2002wt}
between the relative couplings of neutral Higgs particles to the
gauge bosons (\ref{Eq:WW}) and fermions (\ref{Eq:chi-ud}) ({\sl
basic relative couplings}). Since such couplings can be treated as
measurable quantities,  relations between them  are especially
useful in phenomenological analyses.

Let us remind that in these relations we use the quantity
$\tan\beta_{II}$ which coincides with the ratio $v_2/v_1$ only for
a Model II Lagrangian (and has no this simple sense for other
forms of Lagrangian). It is described via the basic relative
couplings for $h_i$ as
\begin{equation}
\!\!\tan^2\beta_{II}\!=
\!{\fr{(\chi_V^{(i)}\!-\!\chi_d^{(i)})^*}{\chi_u^{(i)}\!-\!\chi_V^{(i)}}}
\!=\! {\fr{1\!-\!|\chi_d^{(i)}|^2}{|\chi_u^{(i)}|^2\!-\!1}}= \!
\fr{\Im \chi_d^{(i)}}{\Im \chi_u^{(i)}}\!. \label{Eq:tan-beta}
\end{equation}
Certainly, these expressions hold also for
$h,H,A$, except the last one, which  is
absent for $h,H$.

1. {\em The pattern relation} among the basic relative couplings
holds of {\it each neutral Higgs particle $h_i$}
(in particular also for $h,H,A$ in the case of CP conservation)
\cite{Ginzburg:2001ss,Ginzburg:2002wt}:
 \bear{c}\label{2hdmrel}
(\chi_u^{(i)} +\chi_d^{(i)})\chi_V^{(i)}=1+\chi_u^{(i)}
\chi_d^{(i)}\,,
 \\[1mm]\mbox{ or }\\[1mm]
(\chi_u^{(i)}-\chi_V^{(i)})
(\chi_V^{(i)}-\chi_d^{(i)})=1-(\chi_V^{(i)})^2.\end{array}
\end{equation}

2. {\it A vertical sum rule} for each basic relative coupling
$\chi_j$ for  {\it all three neutral Higgs bosons $h_i$ is given
by} \cite{Grzadkowski:1999ye}:
 \be
\sum\limits_{i=1}^{3}(\chi_j^{(i)})^2=1\,\qquad (j=V,d,u)\,.
\label{vsr}
 \ee

For couplings to the gauge bosons this sum rule, written also
above in eq. (\ref{Eq:W}b), takes place independently on a
particular form of the Yukawa interaction.

3. The relations (\ref{Eq:chi-ud}) allow also to write for each
neutral Higgs boson $h_i$ {\em a horizontal  sum rule}
\cite{Grzadkowski:1999ye}:
\begin{equation}
|\chi_u^{(i)}|^2\sin^2\beta_{II}+|\chi_d^{(i)}|^2\cos^2\beta_{II}=1\,.\label{srules}
\end{equation}
These sum rules guarantee that  the cross section to produce each
neutral Higgs boson $h_i$ (or $h,H,A$) of the 2HDM,  in the
processes involving Yukawa interaction, cannot be lower than that
for the SM Higgs boson with the same mass
\cite{Grzadkowski:1999ye}.

4. Besides, the  useful {\it linear relation} follows directly
from Eqs.~(\ref{Eq:WW}), (\ref{Eq:chi-ud}):
 \bear{c}
 \chi_V^{(i)}=
\cos^2\beta_{II} \,\chi_d^{(i)*}+\sin^2\beta_{II}\, \chi_u^{(i)}=\\[3mm]
=\cos^2\beta_{II} \,\chi_d^{(i)}+\sin^2\beta_{II}\, \chi_u^{(i)*}
\Rightarrow\\[3mm]
\left\{\!\begin{array}{c}\chi_V^{(i)}\!=\!\Re\left(\cos^2\beta_{II}
\chi_d^{(i)}\!+\!\sin^2\beta_{II}
 \chi_u^{(i)}\right), \\[2mm]
 \Im\left(\cos^2\beta_{II} \chi_d^{(i)}-\sin^2\beta_{II}
 \chi_u^{(i)}\right)=0.
\end{array}\right. \end{array}\label{reimchi}
 \ee
5. {\it  The relation for CP violated parts of Yukawa couplings}:
is obtained by exclusion of $\beta_{II}$ from the equations
\eqref{srules}, \eqref{reimchi}
  \be
 (1-|\chi_d^{(i)}|^2)\,\Im\chi_u^{(i)}
  +(1-|\chi_u^{(i)}|^2)\,\Im\chi_d^{(i)}=0\,.
  \label{newrelchi}  \ee

\subsubsection{Some applications.}
Let us remind that the relative couplings to quarks are generally
complex in contrast to the couplings to gauge bosons. For $h_i$
 (or $h,H,A$) we found the following results.
\begin{Itemize}
\item From (\ref{srules}) we get
 \bear{c}
|\chi^{(i)}_u|\gg 1 \;\Rightarrow\;\tan\beta_{II}\ll 1\,;\\
|\chi^{(i)}_d|\gg 1 \;\Rightarrow\;\tan\beta_{II}\gg
1\,.\label{lstan}
 \end{array}\ee
\end{Itemize}
It is instructive to consider now consequences of the  relations
(\ref{2hdmrel})--(\ref{vsr}) for the case when some basic relative
couplings of a Higgs boson are close to $\pm 1$.
\begin{Itemize}
\item
 In virtue of (\ref{srules})  we have for moderate $\tan\beta$
 \be
|\chi^{(i)}_u|\approx 1\;\Rightarrow\;|\chi^{(i)}_d|\approx
1.\label{all1}
 \ee
 \end{Itemize}

Note, that if $\tan\beta$ is extremely large or extremely small,
{\it the horizontal sum rule} allows   $|\chi^{(i)}_d|$ to differ
strong from 1 or $|\chi^{(i)}_u|$ to differ strong from 1,
respectively (in agreement with ~(\ref{lstan})).

Taking for definitness the case of $\chi_j^{(2)}\approx \pm 1 $,
we get:

 \begin{Itemize}
 \item
From (\ref{vsr}),
 \bear{c}
\chi^{(2)}_u\approx \pm
1\;\Rightarrow\;\chi^{(1)}_u\approx \pm i\chi^{(3)}_u\,,\\
\chi^{(2)}_d\approx \pm 1\;\Rightarrow\;\chi^{(1)}_d\approx \pm
i\chi^{(3)}_d\,.
\end{array}\label{couplconj}\ee
\item For $\chi_V^{(2)} \sim\pm 1$
 \bear{c}
\mbox{\it if}\qquad \chi_V^{(2)}\approx\pm 1\\[2mm] \Rightarrow
\left\{\begin{array}{cl}
 (a)&\chi_u^{(2)}\approx \chi_V^{(2)}\; \;or\;\;
 \chi_d^{(2)}\approx \chi_V^{(2)}\,,\\[2mm]
 (b)& \chi_u^{(2)}\approx \chi_d^{(2)}\approx\chi_V^{(2)}\,,\\[2mm]
 (c)& \chi_V^{(1)}\approx\chi_V^{(3)}\approx  0\,,\\[2mm]
 (d)& \chi_u^{(1)}\,\chi_d^{(1)}\approx \chi_u^{(3)}\,\chi_d^{(3)}
 \approx -1\,.\end{array}\right.\end{array}\label{gauge2coup}
  \ee
 \end{Itemize}

The property (a) obtained from (4.8b),  means that the coupling of
$h_2$ to at least one fermion type ($u$ or $d$) is close to the
$\chi_V^{(2)}$. The property (b) follows from property (a) and
(\ref{reimchi}), at moderate $\tan\beta$. The fact that the
couplings of Higgs bosons to gauge bosons are real leads, with the
aid of (\ref{vsr}), to the property (c). Taking into account
property (c) and the pattern relation (4.8a) \ \ we obtain
property (d): the product of Yukawa couplings for other Higgs
bosons (not $h_2$) is close to the corresponding product for
pseudoscalar $A$ in the CP conserving case.

Certainly, results analogous to (\ref{couplconj}),
(\ref{gauge2coup}) hold in the cases when $\chi_j^{(1)}\approx \pm
1$ or $\chi_j^{(3)}\approx \pm 1$.

\subsection{Comments on radiative corrections}\label{secRC}

All  results described so far were obtained in the tree
approximation. Let us discuss briefly the stability of relations
\eqref{2hdmrel}--\eqref{newrelchi} among in principle {\it measurable}
parameters of the  model  in respect to the radiative corrections (RC)
(treated mainly as the one--loop effects).

Certainly, the observable quantities should be obtained from the
Lagrangian (and potential) with  RC. Then one can treat the
presented relations \eqref{2hdmrel}--\eqref{newrelchi} as obtained from the
renormalized parameters (the elements of mass-squared matrix $\cal
M$, the v.e.v.'s ratio $\tan\beta$ (\ref{Eq:vevs-tanbeta}) and the
corresponding Euler angles $\alpha_i$ of Eq.~(\ref{Eq:R-angles})).

The approach which we adopt in our analysis is to  deal with {\it
the relative couplings} (\ref{Eq:gj}) -- the ratios of  the
couplings of each neutral Higgs boson $h_i$ to the gauge bosons
$W$ or $Z$ and to quarks or leptons ($j=W,Z,u,d,\ell...$), to the
corresponding SM couplings. We assume that for each such relative
coupling the RC are included in both: the couplings of the 2HDM
(in the numerator) and those of SM (in the denominator). The largest RC to the
Yukawa $\phi \bar{q}q$ couplings are the one--loop QCD corrections
due to the gluon exchange. They are identical in the SM and in the
2HDM. If they are relatively small, they cancel in both ratios
$\chi_u$ and $\chi_d$. The same is valid for purely QED  RC to all
basic couplings as well as for electroweak corrections including
virtual $Z$ or $W$ contributions.

The situation is different for the electroweak corrections
containing Higgs bosons in the loops. They are different in the SM and
 2HDM, moreover their values depend on the parameters of 2HDM. These
type of RC may modify slightly  some relations presented in
sec.~\ref{secYuk}. However, it is naturally to expect that these
RC are small (below 1 \%) except for  some small corners of
parameter space.

A  delicate problem appears in a description of RC for the
physical states after EWSB. The physical Higgs states become
unstable and they have no asymptotic states. Therefore, the
scattering matrix, written in terms of these fields, becomes
non-hermitian. In particular, the mass matrix for Higgs bosons,
obtained from (\ref{Eq:M3by3}) with RC, become non-hermitian. Full
treatment of this problem demands a subtle theoretical analysis.
In practice, we limit ourselves to some reasonable approximations.
The effects of instability can be neglected when these Higgs
bosons are almost stable (their  widths are much smaller than the
masses and  mass splittings). These effects should be taken into
account in the case of the approximate mass degeneracy, i.e. when
some of masses $M_i$ are very close to each other. In such case a
good description of the masses and couplings is given by an
approximation in which a (complex) matrix of polarization
operators is added to the mass matrix (\ref{Eq:M3by3}).

\section{ Constraints for Higgs Lagrangian}\label{secpert}


The parameters of Higgs potential are constrained by three
types of conditions:
\begin{Itemize}
\item positivity (vacuum stability) constraints
 \item minimum constraints
\item tree-level unitarity  and  perturbativity constraints,
\end{Itemize}
which we will discuss below. The positivity and unitarity
constraints were discussed in literature till now only for the
case of a soft $Z_2$ violation, and  the unitarity constraints
only in the CP conserving case. In the same case of soft $Z_2$
violation the latter constraints were extended to the CP
non-conserving case in \cite{unitCP}. Here we present some new
results for the case of hard $Z_2$ symmetry violation (see
\cite{unitCP1}).

\subsection{\bf Positivity (vacuum stability) constraints}

To have a {\sl stable vacuum}, the potential must be positive at
large quasi--classical values of fields $|\phi_k|$ ({\sl
{positivity constraints}}) for an arbitrary direction in the
$(\phi_1,\phi_2)$ plane. These constraints were obtained for the
case of soft $Z_2$ violation (see e.g.\linebreak[4]
\cite{dema,unitCP,Kastening:1992by,Gunion:2002zf}), they are
 \bear{c}
 \lambda_1>0\,, \quad \lambda_2>0,\;\,
\lambda_3+\sqrt{\lambda_1\lambda_2}>0,\\
\lambda_3+\lambda_4-|\lambda_5|+\sqrt{\lambda_1\lambda_2}>0.\end{array}
\label{positiv}
 \end{equation}

To obtain these constraints it is enough to consider only quartic
terms of the potential.\linebreak[4] Let
$(\phi_1^\dagger\phi_1)=x_1\ge 0$, $(\phi_2^\dagger\phi_2)=x_2\ge
0$. Than $(\phi_1^\dagger\phi_2)=\sqrt{x_1x_2}\,ce^{i\alpha}$ with
$|c|\le 1$ (due to Schwartz theorem). The quadratic form
$V(x_1,\,x_2)$ should be positive at large $x_i$ at different $c$
and $\phi$. At $x_2=0$ or $x_1$ we obtain two first conditions. At
$c=0$ the third inequality is derived. At $c=\pm 1$ with variation
of $\alpha$ in respect to the phase of $\lambda_5$ we obtain the
latter constraint.

\subsection{ Minimum constraints}\label{secmin}

The condition for vacuum (\ref{Eq:min-cond}) describes the {\it
extremum} of potential but not obligatory the minimum. The {\it
minimum constraints} are  the conditions ensuring that above
extremum is a  minimum for all directions in $(\phi_1,\phi_2)$
space, except of the Goldstone modes (the physical fields provide
the basis in the coset). This condition is realized if the
mass-matrix squared for the physical fields is positively defined,
which means that its eigenvalues, i.e. the physical mass squared,
are positive: $ M_{h_{1-3}}^2,\;M_{H^\pm}^2>0$. In some applications
the necessary conditions for that:  positivity of all diagonal
elements, principal minors and the determinant of mass-squared matrix
in  different forms \eqref{Eq:M3by3} or \eqref{semidiag}, are
useful (see discussion in sec.~\ref{secvac}).

\subsection{Unitarity and  perturbativity
constraints}\label{secunit}

The quartic terms of Higgs potential ($\lambda_i$) are transformed
to the quartic self--couplings of the physical Higgs bosons. They
lead, in the tree approximation, to the  s--wave Higgs-Higgs and
$W_LW_L$ and  $W_LH$, etc.  scattering amplitudes for different
elastic channels.  These amplitudes should not overcome unitary
limit for this partial wave  -- that is {\it the tree-level
unitarity constraint}.

The unitarity constraint  was obtained first
\cite{Glashow:1977nt} in the frame of minimal SM, with one
Higgs doublet and Higgs potential $V=
(\lambda/2)(\phi^\dagger\phi-v^2/2)^2$ as the condition $3
\lambda\le 8\pi$. In this model the Higgs-boson mass
$M_H=v\sqrt{\lambda}$ and its width $\Gamma_H$, given
mainly by a decay of Higgs boson to the longitudinal
components of gauge bosons $W_L$, $Z_L$ (originated from
the Goldstone components $G^{\pm}$), grow with $M_H$ as
$M_H^3$. Therefore, the unitarity limit corresponds
simultaneously to  the case where $\Gamma_H\approx M_H$, so
that the physical Higgs boson disappears. On the other hand
it is well known that for $\lambda\gtrsim 8\pi$  at
$\sqrt{s}> v\sqrt{\lambda}\gtrsim v\sqrt{8\pi}\approx 1.2$
TeV the Higgs boson self--interaction  become strong, it is
realized as a strong interaction  of $W_L$ and $Z_L$
(appeared as a Goldstone modes of a Higgs  doublet at
EWSB). Therefore, the unitarity limit is a boundary (in
$\lambda$'s space) between two different physical regimes.
Below the unitarity limit we have more or less narrow Higgs
boson with well known properties (and no strong interaction
effects in the Higgs sector). Above the unitarity limit the
Higgs boson disappears as a particle, discussion in terms
of the observable Higgs particle becomes senseless, and the
Higgs sector becomes strongly interacting.

Akeroyd {\it et al.}\ \cite{Akeroyd:2000wc} have derived the
unitarity constraints for the 2HDM without a hard  violation of
$Z_2$ symmetry for the CP conserving case, i.e. for real
$\lambda_{1-5}$. In the  general CP nonconserving case with soft
violation of $Z_2$ symmetry the parameter $\lambda_5$ is complex.
The application of the RPhT \eqref{Eq:rephase} allows to eliminate
phase of $\lambda_5$, coming to the rephasing equivalent
Lagrangian with real $\lambda_5^s\equiv |\lambda_5|$ ($m_{12}^2$
remains complex). Use of this Lagrangian allows to extend the
results presented in \cite{Akeroyd:2000wc} for unitary constraints
to the CP nonconserving case \cite{unitCP}.

In the considered cases the $Z_2$ symmetry is not violated by the
quartic terms of potential. Unitarity constraints are written  in
ref.~\cite{unitCP} as the bounds for the eigenvalues $\Lambda^{Z_2
parity}_{Y\sigma}$ of the high energy Higgs--Higgs scattering
matrix for the different quantum numbers of an initial state:
total hypercharge $Y$, weak isospin $\sigma$ and $Z_2$ parity.
These bounds given separately for the $Z_2$-even ($\phi_1\phi_1$
and $\phi_2\phi_2$) and $Z_2$-odd ($\phi_1\phi_2$) initial states
are as follows:
  \bea
&|\Lambda_{Y\sigma\pm}^{Z_2}|<8\pi\;\qquad\mbox{\it with }
&\nn\\
 &\Lambda_{21\pm}^{even}=\fr{1}{2}\left(\lambda_1+\lambda_2\pm
 \sqrt{(\lambda_1-\lambda_2)^2+4|\lambda_5|^2\,}\;\right),&\nn\\
&  \Lambda_{21}^{odd}=\lambda_3+\lambda_4\,,\quad
\Lambda_{20}^{odd}=\lambda_3-\lambda_4\,,&\nn\\
&\Lambda_{01\pm}^{even}=\fr{1}{2}\left(\lambda_1+\lambda_2\pm
 \sqrt{(\lambda_1-\lambda_2)^2+4\lambda_4^2\,}\right),&\nn\\
& \Lambda_{01\pm}^{odd}=\lambda_3\pm|\lambda_5|\,,&\label{eigen}\\
&\Lambda_{00\pm}^{even}=\fr{3(\lambda_1+\lambda_2)\pm
 \sqrt{9(\lambda_1-\lambda_2)^2+4(2\lambda_3+\lambda_4)^2\,}}{2},&\nn\\
 &\Lambda_{00\pm}^{odd}=\lambda_3+2\lambda_4\pm
 3|\lambda_5|\,.&\nn
 \eea
For real $\lambda_5$ these conditions coincide with those from
\cite{Akeroyd:2000wc}, obtained however without the above
mentioned identification of various contributions.

At small $\nu$ these constraints result in moderately large upper
bound of $600\div 700$ GeV for $M_H$, $M_A$, $M_{H^\pm}$ (see
examples in Table~\ref{Tabset} of sec.~\ref{sectnondecoup}), see
also e.g. \cite{Akeroyd:2000wc} for the CP conserving case. At
large $\nu$, all Higgs bosons except $h_1$ become heavy without
violating of the unitary constraints (\ref{eigen}).

The correspondence between  a violation of the tree-level
unitarity limit and a lack of realization of the Higgs field as a
resonance (a particle), as in the minimal SM, takes place in the
2HDM only in the case when {\sl all} constraints (\ref{eigen}) are
violated simultaneously. In the case when only some of these
constraints are violated the physical picture become more complex.
One can imagine, for example, a situation when some of the Higgs
bosons  are "normal"\ scalars, i.e. their properties can be
estimated perturbatively, while the others interact strongly at
sufficiently high energy. In such case, the unitarity constraints
work differently for different {\it physical} channels, in
particular, for different Higgs bosons.

The {\it perturbativity condition (constraint)} for a validity of
a tree approximation  in the description of some particular
phenomena (e.g. interactions of the lightest Higgs boson $h_1$)
may be less restrictive than the presented above general unitarity
constraints. The explicit form of the perturbativity constraint
should be found,  however this is a subject for a separate
consideration. In particular, the effective parameters of
perturbation theory for the Yukawa interaction is $g^2/(4\pi)^2$.
Therefore, one of the necessary conditions for the smallness of
radiative corrections is $ |g|\ll 4\pi$.

\subsection{ The case of hard $Z_2$ violation}\label{secperthard}

The analysis of the case with hard $Z_2$ violation (i.e. the
potential with $\lambda_{6,7}$ terms) is more complicated. One can
say definitely that the positivity constraints (\ref{positiv}) are
valid for some particular directions of a growth of the
quasi-classic fields $\phi_{1,2}$. Similarly, unitarity
constraints (\ref{eigen}) hold for such transition amplitudes
which don't violate the $Z_2$ symmetry.

For the hard violation of $Z_2$ symmetry one should consider new
directions in the $(\phi_1,\phi_2)$ space which appear  due to
$\lambda_6$, $\lambda_7$ terms and the processes like
$\phi_1\phi_1\to \phi_1\phi_2$, which violate the $Z_2$ symmetry.
Therefore the new positivity and unitarity constraints should
include parameters $\lambda_6$, $\lambda_7$. In any case
conditions \eqref{eigen} are necessary for unitarity
\cite{unitCP1}.

\section{Heavy Higgs bosons in 2HDM}\label{secheavy}


Many analyses of 2HDM assume a SM--like physical picture: the
lightest Higgs boson $h_1$ is  similar to the Higgs boson of the
SM while other Higgs bosons escape observation being too heavy.
Besides, many authors assume {\it in addition} that  masses of
other Higgs bosons $M$ are close to the scale of new physics,
$M\sim \Lambda$, and  that the theory should possess an explicit
{\bf decoupling property}\fn{Generally, this property is important
feature of any  consistent theory describing phenomena at some
distances (energies), that is an independence of its predictions
from the dynamics at smaller distances, described by some mass
scale $M$ \cite{Appelquist:tg}. },  i.e. {\it the correct
description of the observable phenomena must be valid
for the (unphysical)  limit $M  \to\infty$}
\cite{Kanemura:1996eq,Ciafaloni:1996ur,Malinsky:2002mq,Kanemura:2002cc,
Gunion:2002zf}. However, the 2HDM allows also for another
realization of the mentioned SM--like physical picture.

Looking on formulae from sec.~\ref{secphysHig} we see that the
large masses of Higgs particles may arise from large parameters
$\nu$ or $\lambda's$, or both. Obviously, large values of
$\lambda's$ may be in conflict with unitarity constraints, which
is not the case for large $\nu$. Below we discuss these two very
distinct  sources of large masses, and their different
phenomenological consequences.

\subsection{ Decoupling of heavy Higgs bosons}\label{secdecoupl}

In 2HDM  the decoupling case  corresponds to
 \be
 \nu\gg |\lambda_i| \,.\label{declim}
 \ee
That means that it can not be realized for the exact $Z_2$
symmetry. In the decoupling case equations for masses and mixing
angles $\alpha_i$ (\ref{Eq:mHh}) simplify. First we find, with
accuracy up to the $\lambda/\nu$ terms,  masses of the subsidiary
Higgs states obtained at the first stage of diagonalization
(\ref{Eq:diagonal-2times2}-\ref{mixCPgen}). From
eqs.~(\ref{massevendiag}) we derive
  \bea
&\fr{M_h^2}{v^2}=\underbrace{c_\beta^4\lambda_1+s_\beta^4\lambda_2+
2s_\beta^2c_\beta^2\lambda_{345}}_{soft}+
\underbrace{4s_\beta^2c_\beta^2\Re\lambda_{67}}_{hard},&\nn\\
&\fr{M_H^2}{v^2}=\underbrace{\nu+
s_\beta^2c_\beta^2(\lambda_1+\lambda_2-2\lambda_{345})}_{soft}&\label{decmass}\\
&-\underbrace{[(c_\beta^2-s_\beta^2)\Re\tilde{\lambda}_{67}+(4s_\beta^2c_\beta^2-1/2)
\Re\lambda_{67} ]}_{hard}.&\nn
 \eea

In the proper decoupling limit $\nu\to\infty$ we have
$\beta-\alpha\to \pi/2$. It is useful to characterize a deviation
from  this limiting value by a parameter\linebreak[4]
$\Delta_{\beta\alpha}=\pi/2-(\beta-\alpha)$. Using
$s_{2\beta}=\sin 2\beta$,\linebreak[4] $c_{2\beta}=\cos 2\beta,$
we get from the second line of eq.~(\ref{Eq:mHh}):
  \bear{c}
 \Delta_{\beta\alpha}=-\fr{L_as_{2\beta}}{2\nu}\,,\\
 L_a=
  \underbrace{
s_\beta^2\lambda_2-c_\beta^2\lambda_1
+c_{2\beta}\lambda_{345}}_{soft}\\ + \underbrace{\Re( 2c_{2\beta}
\lambda_{67} -\tilde{\lambda}_{67})}_{hard}.
\end{array}\label{decmass1}
 \ee

The subsequent complete diagonalization, described in
sec.~\ref{refcompl}, is simplified by condition (\ref{declim}). We
get the following results.

\subsubsection{\bm The lightest Higgs boson $h_1$.}

The eq-s \eqref{mixCPgen}, (\ref{newlamconstr})
show that under the condition (\ref{declim}) the element
$M'_{13}$\ of the matrix (\ref{Eq:diagonal-2times2}), responsible
for the mixing of scalar $h$ with $A$, is small as compared to the
mass difference $M_A^2-M_h^2\approx \nu v^2$. Therefore, the state
$h_1$ is very close to $h$. The mixing angle $\alpha_2$,
describing the CP--odd admixture in this state, is given by
$s_2\sim |\lambda|/\nu(\ll 1)$ (\ref{weakCPsi}). The shift of the
mass of $h_1$ from the $M_h$ value (\ref{decmass}) is given by
Eq.~(\ref{weakCPssi}), i.e. $M_1^2-M_h^2\sim |\lambda|/\nu$, and
can be neglected.

Since for $\nu \to \infty$ the $\Delta_{\beta\alpha} \to 0 $ the
scalar $h_1$ couples to the gauge bosons and  to the quarks and
leptons in the Model II as in the SM (with accuracy
$|\lambda|/\nu$) even for the general CP non-conserving case.
Besides, $h_1$ practically decouple from $H^\pm$, since the
quantity $\chi^{(1)}_{H^\pm} \sim{\cal O}(|\lambda|/\nu)  $
(\ref{b2d3}).

\subsubsection{\bm Higgs bosons $h_2,h_3$ and $H^{\pm}$.}

The eqs.~ (\ref{Eq:mch}), (\ref{Eq:M3by3}c), (\ref{decmass}) show
that
 \bes\be
M_{H^\pm}^2\approx M_A^2\approx M_H^2=v^2{\nu}\left[1+{\cal
O}\left(\dfrac{|\lambda|}{\nu}\right)\right],
 \ee
i.e. $H^\pm$, $H$ and $A$ are very heavy and almost degenerate in
masses, and similarly for $h_2$ and $h_3$
 \be
M_{H^\pm}^2 \approx M_2^2\approx M_3^2\approx
v^2{\nu}\left[1+{\cal
O}\left(\dfrac{|\lambda|}{\nu}\right)\right].
 \ee\ees
That is one of the reasons to consider the condition of the
decoupling regime (\ref{declim}) in the form, used e.g. in
ref.~\cite{Gunion:2002zf},
 \be
 M_A^2\gg |\lambda|v^2 \,.\label{declimHab}
 \ee

In the considered case the CP violating mixing between $H$ and $A$
can be strong, i.e.  mixing angle $\alpha_3$ given by
eq.~(\ref{tanalph3}), can be large as it was discussed in
sec.~\ref{secsomecase}.

Since $\chi_V^{(h)}\approx 1$, the coupling of $H$ to gauge bosons
is very small, while $A$ does not couple to gauge bosons
(Table~\ref{couptab}).  With mixing between $H$ and $\,A$ states given
by angle $\alpha_3$, we have
 \bes \label{extradecoupl}\bear{c}
\chi_V^{(H)}=\cos(\beta-\alpha)\approx \Delta_{\beta
\alpha}\,,\;\;\chi_V^{(A)}=0\Rightarrow\\ \chi_V^{(2)}\approx
-\cos\alpha_3 \,\Delta_{\beta \alpha},\;\, \chi_V^{(3)}\approx
\sin\alpha_3 \,\Delta_{\beta \alpha}\end{array}.
 \ee

Besides, the couplings of $H$ and $A$ to the $u$-type
fermions coincide in their modules (see Table~\ref{couptab})
(and the same is valid for $d$-type fermions and charged
leptons), so that also the corresponding couplings of $h_{2,3}$
have equal modules, while their phases, related to the CP
violation in the $(\bar{u}h_{2,3}u)$ and $(\bar{d}h_{2,3}d)$
vertices, are given by the mixing angle $\alpha_3$. Using
eqs.~(\ref{Eq:chi-ud}) and (\ref{Eq:R-angles}) we obtain
 \bear{c}
  \chi^{(2)}_u=i \chi^{(3)}_u=\cot\beta\,e^{-i\alpha_3},\\
 \chi^{(2)}_d=-i
 \chi^{(3)}_d=-\tan\beta\,e^{i\alpha_3}.\end{array}\label{decoupYuk}
  \ee

The corresponding Higgs decay widths are given mainly by the
fermionic contributions,
 \bear{c}
\Gamma_H\approx \Gamma_A\approx
\Gamma_2\approx\Gamma_3\\=\fr{3}{16\pi}\cot^2\beta
 \left[1+
 \fr{m_b^2}{m_t^2}\tan^4\beta\right]M_H\\ \mbox{\it\  with }\;
\fr{\Gamma_A-\Gamma_H}{\Gamma_H}\sim \fr{m_t}{M_H}
 .\end{array}\label{decwidths}
 \ee \ees
(Here we took into account that $v^2/m_t^2\approx 2$.) The gauge
boson contributions to these widths are negligibly small ($\sim
L_a^2/\nu$). Therefore, we have\linebreak[4]
$|M_H^2-M_A|^2\lesssim \Gamma_AM_A,\,\Gamma_HM_H$ at very large
$\nu$. In this case the equations for $\alpha_3$, $M_{2,3}$ and
$\Gamma_{2,3}$ become more complex, since they  include shift of
the $A,\,H$ poles due to their proper widths (\ref{decwidths}).
The obtained  mass matrix become non-hermitian, therefore,  the
mixing angle $\alpha_3$ become complex, and states $h_2$, $h_3$
become non-orthogonal. It is seen from the  eq.~(\ref{tanalph}),
corrected for these effects:
 \be
\tan 2\alpha_3\approx
\fr{2M'_{23}}{M_A^2\!-\!M_H^2\!-\!i(M_A\Gamma_A\!-\!M_H\Gamma_H)}
\label{corralph}
 \ee
(see \cite{Pilaf97,Jan}  for more details of mixing). In this case
eq-s. \eqref{decoupYuk} are modified.

Note that with this strong overlapping of states  experimental
distinguishing of states $h_2$, $h_3$ may be difficult. The
visible effects of CP violation in the fermion interaction (like
spin correlations, etc.) will be very similar in two quite
different cases: of a true CP violation and of a strong
overlapping of $H$ and $A$ states without the CP violating mixing.

\subsection{Heavy Higgs bosons  without decoupling}\label{sectnondecoup}

The option,  where except of one neutral Higgs boson $h_1$ (or
$h$), all other Higgs bosons are reasonable heavy, can also be
realized in 2HDM for a relatively small $\nu$, i.e. beyond the
decoupling limit. In this case possible masses of heavy Higgses
are bounded from above by the unitarity constraints for
$\lambda_i$, discussed in  sect.~\ref{secunit}. These constraints
obtained for the CP conserving case \cite{Akeroyd:2000wc} can be
generally stronger in the case of CP violation, since the
constraints (\ref{eigen}) put limit on parameter $|\lambda_5|$
while formulae for masses contain solely $\Re\lambda_5$. In the
Table~\ref{Tabset} we present some particular examples of sets of
parameters of the potential for light $h$ (mass 120 GeV) and heavy
$H$, $H^\pm$ for a non-decoupling case (small $\nu$) and
satisfying unitarity constraints (\ref{eigen}).

The first three lines contain  sets of parameters $\lambda_i$ and
$\nu$ for the case without CP violation with reasonably heavy $H$,
$H^\pm$, $A$. One sees that  these masses can be obtained for very
large or very small $\tan\beta$ and reasonably small $\nu\approx
(M_h/v)^2$, as well as for $\tan\beta\approx 1$ with $\nu\approx
0$.

The fourth line of the Table~\ref{Tabset} presents an example of
the {\it natural set of parameters} (see below),  with heavy $H$
and $H^\pm$ in the weak CP violation case. Since here mixing
angles $\alpha_2$, $\alpha_3$ are small, the physical states
$h_1$, $h_2$, $h_3$ are close to the states $h$, $-H$ and $A$,
existing in the CP conserved case.

In  the considered non-decoupling case couplings  of the
lightest Higgs boson to the gauge bosons, quarks  and
leptons  can  be either close to the corresponding SM
values (as in the decoupling case) or far from these
values. The case when all basic couplings of  the lightest
Higgs boson are close to those of SM  Higgs boson is
discussed in detail in paper \cite{GK2}, see
 also
\cite{Ginzburg:2001ss,Ginzburg:2002wt}. Note,  that even in such
case some non-decoupling effects due to heavy Higgs bosons may
appear, e.g. $\chi_{H^\pm}^{(1)} \sim 1$, in contrast to the
decoupling limit, discussed in sec.~\ref{secdecoupl}, where
$\chi^{(1)}_{H^\pm}\sim 0$. It is worth noticing, that $\nu$
parameter can be negative, which is not possible in the decoupling
limit.
 \begin{widetext}\begin{center}\begin{table}[hbt]
 \caption{\it Sets of parameters of potential  for
light $h$ (mass 120 GeV) and heavy $H$, $H^\pm$ satisfying
unitarity constraints  in the nondecoupling case.}
\label{Tabset}
\begin{center}
\begin{tabular}{|c|c|c|c|c|c|c||c|c|c|c||c|c|}\hline
$\tan\beta$&$\lambda_1$&$\lambda_2$&$\lambda_3$&$\lambda_4$&
 $\lambda_5$&$\nu$ &$M_h$&$M_H$&$M_A$&$M_{H^\pm}$&$s_2$&$s_3$
 \\\hline
50&1&6&5.5&-6&-6&0.24&120&600&600&600&-&-\\
0.02&6&1&5.5&-6&-6&0.24&120&600&600&600&-&-\\
1&6.25&6.25&6.25&-6&-6&0&120&600&600&600&-&-\\\hline

 10&4&8&4.4&-9&$-0.5+0.3i$&0.24&120&700&206&556&0.09&0.02\\\hline
\end{tabular}
\end{center}
 \end{table}\end{center}\end{widetext}

\subsection{A natural set of parameters of  2HDM}

\bu It is natural to consider 2HDM  as low energy approximation of
some more general theory operating at smaller distances. In such
theory  fields $\phi_1$ and $\phi_2$ should differ in some quantum
numbers which cannot be seen at our relatively large distances
(like in MSSM). Therefore, it is naturally to assume that the 2HDM
describing physical reality, allows  an existence among the
reparametrization equivalent Lagrangians the one in which fields
$\phi_k$ don't mix at small distances (mixed kinetic term does not
appear). That is the 2HDM with exact or softly violated $Z_2$
symmetry. We assume such choice in this section.

\bu Besides, it is naturally to assume that the CP symmetry in
Higgs sector is violated only weakly at least for the lightest
Higgs boson $h_1$. This assumption together with rephasing
invariance offers the basis for the selection of the {\it natural
set of parameters of 2HDM}.

The  eq.~(\ref{Eq:diagonal-2times2}) shows that the CP symmetry
for the lightest Higgs boson is violated weakly if and only if
$|M'_{13}|\ll |M_A^2-M_h^2|$. In view of \eqref{mixCPgen}, for
{\it  the real vacuum Lagrangian} at $\beta+\alpha\neq \pi/2$ this
condition can be rewritten in the form
 \be
v^2|\Im m_{12}^2|\ll v_1v_2|M_A^2-M_h^2|\,.\label{nat1} \ee

For all other rephasing equivalent Lagrangians the condition
corresponding to the equation (\ref{nat1}) contains both $\Im
m_{12}^2$ and $\Re m_{12}^2$. Therefore, for {\it the natural set
of parameters of 2HDM} we require that both $|\Im
m_{12}^2|(v^2/v_1v_2)$ and $| \Re m_{12}^2|(v^2/v_1v_2)$ are small
for all rephasing equivalent Lagrangians. In virtue of
(\ref{nudeldef}), (\ref{newlamconstr}) in the case of soft
violation of $Z_2$ symmetry the same requirements is transmitted
to  $\Im\lambda_5$ and  $\Re\lambda_5$. Therefore, we define {\it
a natural set of parameters} as follows
 \be
 |\nu|,\;|\lambda_5|\ll
|\lambda_{1-4}|\;.
 \label{nat2}
 \ee
Contrary, in the decoupling case, the term  $m_{12}^2$ has the
{\it unnatural property} $\Re m_{12}^2\gg |\Im
m_{12}^2|$. From this point of view {\it the decoupling case of
2HDM (\ref{declim}) is unnatural}.

For the natural set of parameters of 2HDM the breaking of the
$Z_2$ symmetry is  governed by a small parameter $\nu$. Due to the
existence of a limit when $Z_2$ symmetry holds, a small soft $Z_2$
violation in the Higgs Lagrangian and the Yukawa interaction
remains small also beyond the tree level. In this respect we use
term {\it natural} in the same sense as in ref.~\cite{'tHooft:xb}.
(Note that also non-diagonal Yukawa coupling matrices $\Gamma_1$
and $\Delta_2$ (leading to FCNC) are unnatural in this very
sense).

In accordance with Eq.~(\ref{Eq:M3by3}),  for the natural set of
parameters also $M_A$ cannot be too large (see
Table~\ref{Tabset}). This opportunity is not ruled out by data,
see for CP conserving case e.~g. \cite{taukt}.

\bu {\bf Yukawa sector}. In the case of true hard violation of
$Z_2$ symmetry the Yukawa sector cannot be described by simple
model of type I or II, i.e. models like Model III should be
realized. However  in such models the FCNC effects (and CP
violation in the Higgs sector) are naturally large. That is an
additional reason why the natural set of parameters of 2HDM
corresponds to the case of exact or softly violated $Z_2$
symmetry with the  Model II or Model I for Yukawa interaction.

\section{Summary and discussion of results}\label{secsum}

In this paper we analyse various aspects of the
two--Higgs--doublet extension of the SM from point of view of
its symmetries. We critically discuss the standard formulations as
well as applications of  the 2HDM. Let us describe our approach
and summarize main results and observations presented in the
paper.

\bu At the beginning we stress that the CP violation can be
implemented in a model in a few different ways. In this
paper we consider mainly the CP violation govern by complex
parameters of the Higgs Lagrangian. However, there are
other ways of implementation of CP violation. For example
the one, mentioned in sec.~\ref{secYUkgen}, which relies on
complex elements of the Yukawa matrices. Another way, used
in fact in many analyses of MSSM, is based on the CP
nonconservation in the couplings of Higgs bosons to
superpartners. The renormalizability demands to add in such
cases the CP violating terms also in the Higgs Lagrangian.

$\bullet$ In the analysis of symmetry properties of the model we
introduce {\it  the 16-dimensional space of Higgs Lagrangians with
coordinates given by the Lagrangian parameters}. Within this space
there is the 3-dimensional subspace -- {\it the reparametrization
equivalent subspace}, formed by Lagrangians which can be obtained
from a chosen one by the reparametrization transformation RPaT's
(\ref{newlagr}). All the Lagrangians from this subspace describe
the same physical reality (a reparametrization invariance).
Different properties of the physical model can either be explicit or
hidden for  the different
 Lagrangians in the mentioned reparametrization equivalent
subspace. Accordingly, different families
of these Lagrangians are suitable for the study of different
properties of the model. Obviously, all measurable quantities
characterizing a system (like the coupling constants and masses)
are reparametrization invariant while many other parameters of
theory (like $\tan\beta$) are reparametrization dependent.

Certainly the concept of the reparametrization invariance, etc.
can be easily generalized to a description of other models, with
e.g. with 3 or more Higgs doublets sector, etc. and for
description of Yukawa interactions.

The  reparametrization equivalent space is naturally sliced to
{\it the rephasing equivalent subspaces}, which are described by
transformations (\ref{reparam}) with $\theta=0$ (the RPhT's)
represented by vertical strips in FIG.~1. One can characterize
these subspaces e.g. by the value of ratio of v.e.v.'s $tan\beta$.

The CP violation in Higgs sector means that the physical
neutral Higgs bosons have no definite  CP parity. The
necessary condition for such CP violation is that some of
coefficients of the Higgs Lagrangian are complex. However,
complex parameters can appear also in the CP conserving
case if not pertinent form of Lagrangian is chosen. We
found a specific, {\it real vacuum form of a Lagrangian} in
which complexity of the parameters of Higgs Lagrangian
becomes a sufficient condition for the CP violation in
Higgs sector. For the arbitrary form of Lagrangian we give
a simple necessary and sufficient condition for the CP
violation in the Higgs sector \eqref{newcond}. This
condition is  simpler to use than a similar
condition written via IRpaT in ref.~\cite{GH05}.

Some  authors  consider also the case when basic Higgs
Lagrangian give no  CP  violation  in  Higgs sector but
this violation appears through the Yukawa  interaction. The
series of combination of Higgs self-couplings and  Yukawa
couplings  form reparametrization independent invariants,
describing condition for the CP violation. Note that in
this case loop corrections from  the Yukawa interaction
produce  CP violated  terms in the Higgs Lagrangian. From
general renormalizability such terms must be included in
the basic Higgs Lagrangian and our simple criteria for CP
violation (3.16) seem to be sufficient.

\bu The 2HDM provides mechanism of the EWSB which allows for
potentially large CP violation and FCNC effects. These phenomena
are controlled to a large extent by the  $Z_2$ symmetry under
transformation (\ref{Eq:Z2-symmetry}) of the Lagrangian and
various degree of its violation. If the $Z_2$ invariance holds,
then the considered doublets of scalar fields $\phi_{1,2}$ are the
true fundamental basic fields before EWSB. The  soft violation of
$Z_2$ symmetry is given by the mixed mass term $\sim m_{12}^2$ in
the Higgs potential. In this case  two doublets $\phi_{1,2}$ mix
near EWSB scale but they don't mix at sufficiently small
distances. The RPaT converts such Higgs Lagrangian, ${\cal L}_s$,
to the form  with terms typical for a hard violation of  the $Z_2$
symmetry ({\em a hidden soft $Z_2$ violation form of
Lagrangian}). However, in this case the parameters of the Higgs
potential are interrelated as it is given by eq.~(\ref{softlagr})
(see also eqs.~(\ref{hidsoftconstr})). It prevents an appearance
of a running coefficient at the  mixed kinetic term.

In the case of true hard violation of $Z_2$ symmetry  even the
discussion of  Higgs potential alone is incomplete, since it is
necessary to consider more general Higgs Lagrangian with the  mixed
kinetic term. The coefficient of  this mixed term of Lagrangian
$\vak$ (\ref{kinterm}) generally runs due to the loop corrections.
At some fixed distance (renormalization scale) the kinetic part of
the Lagrangian can be removed by diagonalization like
(\ref{diagkap}) but this term is restored at other distances
(renormalization scales) due to the loop corrections from hard terms
of the Higgs potential. We did not find  a fully consistent
formulation of 2HDM in the case when the  mixed kinetic term is
present. We argue, that due to the mentioned relation to the
phenomena at small distances, the case with soft violation of
$Z_2$ symmetry looks much more attractive and natural.

In our calculation  we keep separately contributions of soft and
hard violation of $Z_2$ symmetry.  Nevertheless, our discussion of
a hard violation of $Z_2$ symmetry is as incomplete as all other
existing analyses, since
effects related to the running coefficient of the mixed kinetic
term should be analysed in addition.

\bu The EWSB appears at the minimization of a Higgs potential
giving the vacuum expectation values for two scalar fields,
$\phi_1$ and $\phi_2$.For some set of parameters of Lagrangian
these v.e.v.'s describe standard (neutral) vacuum. Generally,
phases of these v.e.v.'s differ from each other. However, this
phase difference can be eliminated by a suitable rephasing
transformation giving  mentioned above the real vacuum Lagrangian.
We prefer to express the mass coefficients of Higgs potential via
$v_{1,2}$ and the free dimensionless parameter $\nu\propto \Re
m_{12}^2$, (\ref{nudeldef}). We use in our analysis such form  of
Lagrangian.

For other set of parameters of Lagrangian the condition of
minimum of the potential defines also  exotic "charged vacuum"
with vacuum energy  larger than that for the standard vacuum
\cite{Diaz-Cruz:1992uw}, \cite{chargebr}.

Some physical model ("physical reality") is described by
many reparametrization equivalent Lagrangians. In contrary,
the description of Lagrangian in terms of the observable
Higgs fields $h_i$ is unique (reparametrization invariant).
For the neutral Higgs sector the transition from fields
$\phi_1$ and $\phi_2$ to the basis of observable Higgs
bosons is rather complicated. We have performed this in two
steps. First, we diagonalize the CP-even part of the
mass-squared matrix. For the Lagrangian in a real vacuum
form this step is identical to the one used in the CP
conserving case. It allows to describe the general CP
violating case in terms of the well known states $h$, $H$
and $A$ treated now as the subsidiary states (i.e. having
no direct physical meaning). Using these states it becomes
evident that the existence of complex coefficients in the
Higgs potential in  a real vacuum form is necessary and
sufficient condition for the CP violation in the Higgs
sector. Our procedure allows  to analyze easily various
important cases when one of neutral Higgs boson is almost
the CP-even one, while two other neutral Higgs bosons
strongly mix, i.e. CP symmetry can be strongly broken in
the processes with exchange of these Higgs bosons.

\bu Considering the Yukawa interactions we note that  for a case
of true hard violation of $Z_2$ symmetry the most general form
of this interaction (e.g. Model III) should be implemented.
However, we limit ourselves to models in which each fermion
isosinglet couples to only one Higgs field and discuss the flavor
structure of such couplings.  We consider in detail the
Model II. We assume that  the Model II Lagrangian
family coincides with mentioned above family of Lagrangians with
explicit softly violated $Z_2$ symmetry. We prefer to use the
Lagrangian in the form which corresponds to the  real vacuum,
Model II and exact or softly violated $Z_2$ symmetry in the
potential.

\bu In this paper we extend our approach introduced  earlier for
the CP conserving case in \cite{Ginzburg:2001ss,Ginzburg:2002wt}
to the analysis of  the CP nonconserving case. This approach
relies on using the measurable (in principle) Higgs boson masses
and basic relative couplings (\ref{Eq:gj}) plus parameter $\nu$
(\ref{nudeldef}) instead of variety of parameters $\lambda$ and
mixing angles $\alpha_i$, $\beta$. This way phenomenological
analyses become more transparent.

We present a series of relations between different relative
couplings of each Higgs boson,
(\ref{2hdmrel})--(\ref{newrelchi}). Among these relations
there are well known sum rules, the  pattern relation
(obtained by us in \cite{Ginzburg:2001ss} for the CP
conserving case and in \cite{Ginzburg:2002wt} for the CP
violation),  new linear relations and their
combinations. Eq.~(\ref{Eq:tan-beta}) represents the
formulae which allows  to determine the quantity
$\tan\beta$ for the Lagrangian in Model II form,
$\tan\beta_{II}$.

Using these relations we obtained various useful relations among
couplings of Higgs bosons to quarks and gauge bosons in the case
when some of these couplings (or their absolute values) are close
to the corresponding values in the SM
(\ref{lstan})--(\ref{gauge2coup}).

As the mentioned relations between relative couplings are of great
phenomenological importance it was crucial to check how the
radiative corrections influence them; we argue that radiative
corrections  change only weakly the considered relations.

\bu Next we combine and discuss different types of constraints on
the parameters of the Higgs potential (the positivity condition or
-- in other words --  the vacuum  stability condition at large
quasiclassical values of $\phi_k$, the existence of a minimum, the
tree-level unitarity constraint from the Higgs-- Higgs scattering
matrix) both in the CP conserving and CP violating cases. Some of
them were known till now only in the CP conserving case. All known
results  were obtained for the case of soft violation of $Z_2$
symmetry only. We ascertained that some of these results are valid
also in the case of hard violation of $Z_2$ symmetry, as a part of
more general system of constraints.

\bu We  perform the detailed discussion on an opportunity that in
the 2HDM there is one light Higgs boson, while  others are much
heavier, so that they can escape observation. As it was already
claimed in \cite{Ginzburg:2002wt},\cite{Gunion:2002zf} such
situation can be realized in the  different regions of $\nu$. At
$\nu\gg |\lambda_i|$ we have decoupling case in which the lightest
Higgs bososn $h_1$ is  very similar to the SM Higgs boson, while
other Higgs bosons except $h_1$ are very heavy and almost
degenerate in masses. We found simple expressions  for their
couplings which hold for  a possible strong CP violating mixing
among them (\ref{extradecoupl}).

At small $\nu$, the  reasonably heavy  Higgs bosons, lighter
however than $\sim$ 600 GeV,  may appear without violation of
unitarity constraints. This small $\nu$ option looks more natural
from point of view  of the rephasing invariance. Here one can
expect some non-decoupling effects due to the heavy Higgs bosons
\cite{Ginzburg:1999fb, Ginzburg:2001ss,Ginzburg:2002wt}. The
detailed analysis of various SM-like realizations and some
non-decoupling effects is presented in \cite{GK2}.

\bu In the Appendix we present for completeness, a whole set of
self--couplings of physical Higgs bosons in the general CP
violating case. Besides, we present  simple formulae for the CP
conserving, soft $Z_2$ violating case. In addition to the well
known forms of these couplings,  for the case when the Yukawa
interaction is described by Model II, we express all  trilinear
couplings via the Higgs masses and their relative couplings to the
gauge bosons and fermions of the physical Higgs bosons entering
corresponding vertex, and  the parameter $\nu$.

\begin{acknowledgments} Some of results presented here were obtained
together with Per Osland \cite{Ginzburg:1999fb},
\cite{Ginzburg:2001ss, Ginzburg:2002wt}  and we are very grateful
to him for a fruitful collaboration. We express our gratitude  to
A. Djouadi, J. Gunion, H. Haber and M. Spira for discussions of
decoupling in the 2HDM and the MSSM, and to P.~Chankowski,  B.
Grz\c{a}dkowski, W.~Hollik, I.~Ivanov,  M.~Dubinin, M.~Dolgopolov,
J.~Kalinowski, R.~Nevzorov, A. Pilaftsis, J.~Wudka, P.~Zerwas for
various valuable discussions. This research has been supported by
grants RFBR 05-02-16211, NSh-2339.2003.2 in Russia,  European
Commission -5TH Framework contract HPRN-CT-2000-00149 (Physics in
Collision), by Polish Committee for Scientific Research Grant No.
~1~P03B~040~26 and 115/E-343/SPB/DESY/P-03/DWM517/2003-2005.
\end{acknowledgments}

\begin{widetext}

\appendix
\renewcommand{\thesection}{A.}
\section*{Appendix. Higgs self-couplings}
\label{sectrilin}

The fact that the charged Higgs field $H^\pm$ and axial field $A$
are expressed via fields $\phi_k$ and angle $\beta$ allows to
obtain Higgs self--couplings  in two-step procedure. At the first
step we come to basis $\eta_1$, $\eta_2$, $H^\pm$, $A$
(\ref{videf}) -- (\ref{physfields}). Than we perform
transformation (\ref{Eq:R-angles}). It results in appearance of
suitable number of matrix elements $R_{ij}$ in addition to a
factors $b_A$ or $a_A$ (eqs. \ref{Eq:charged-coupl},
\ref{Eq:neutral-3-4-coupl}), which are are expressed via couplings
$\lambda_i$, given in the real vacuum form, and  mixing angle
$\beta$. In the equations below the symbol $\ast$ denotes a sum
over permutations over summation indices $i,\,j$, etc. (giving
factors of $n!$ for $n$ identical indices).

For the couplings involving the charged Higgs bosons we have
 \bea
&g_{h_iH^+H^-}=v\sum_{m=1,2,3}R_{im}\,b_m,\quad
g_{h_ih_jH^+H^-}=\sum_{m\le n=1,2,3}^\ast
R_{i'm}R_{j'n}\,\,b_{mn},&\nn\\
&g_{H^+H^-H^+H^-}=2[\sin^4\beta\lambda_1+\cos^4\beta\lambda_2
+2\cos^2\beta\sin^2\beta\Re\lambda_{345}&\label{Eq:charged-coupl}\\
&-4\cos\beta\sin\beta(\sin^2\beta\Re\lambda_6+
\cos^2\beta\Re\lambda_7)].&\nn
 \eea

For the couplings among the neutrals we have
 \bear{c} \label{Eq:neutral-3-4-coupl}
g_{h_ih_jh_k} =\;v\sum^\ast\, R_{i'm}R_{j'n}R_{k'o}\,a_{mno},
\quad g_{h_ih_jh_kh_l} =\sum^\ast R_{i'm}R_{j'n}R_{k'o}R_{l'p}\,
a_{mnop}
 \end{array}\ee
with $m\le n\le o\le p=1,2,3$.

The coefficients $b_m$, $b_{mn}$, $a_{mno}$, $a_{mnop}$ presented
below agree with the corresponding results of
\cite{Choi:1999uk,Carena:2002bb,Akhmetzyanova:2004cy}.

\subsection{General formulae}

\subsubsection*{Trilinear couplings}
The trilinear couplings involving the charged Higgs bosons, are
given by Eq.~(\ref{Eq:charged-coupl}), with
\begin{align}
b_1&=\cos\beta\{\sin^2\beta(\lambda_1-\lambda_{345}) +\lambda_3
+\cos\beta\sin\beta[(\tan^2\beta-2)\Re{\lambda}_6+\Re{\lambda}_7]\},
\nonumber \\
b_2&=\sin\beta\{\cos^2\beta(\lambda_2-\lambda_{345}) +\lambda_3
+\cos\beta\sin\beta[\Re{\lambda}_6+(\cot^2\beta-2)\Re{\lambda}_7]\},
\nonumber \\
b_3&=\cos\beta\sin\beta\,\Im{\lambda}_5-\sin^2\beta\,\Im{\lambda}_6
-\cos^2\beta\,\Im{\lambda}_7.
\end{align}

The trilinear couplings among neutral Higgs fields are given by
Eq.~(\ref{Eq:neutral-3-4-coupl}), where:
 \bear{c}
a_{111}=\fr{1}{2}(\cos\beta\lambda_1+\sin\beta\Re\lambda_6),\quad
a_{112}=\fr{1}{2}(\sin\beta\Re\lambda_{345}+3\cos\beta\Re\lambda_6),
 \\
a_{113}=-\fr{1}{2}
[\cos\beta\sin\beta\Im\lambda_5+(1+2\cos^2\beta)\Im\lambda_6],
\quad
a_{122}=\fr{1}{2}(\cos\beta\Re\lambda_{345}+3\sin\beta\Re\lambda_7),
\\
a_{123}=-\Im\lambda_5-\cos\beta\sin\beta(\Im\lambda_6+\Im\lambda_7),
 \\
a_{133}=\fr{1}{2}
      \{\cos\beta(\sin^2\beta\lambda_1+\cos^2\beta\Re\lambda_{345}
                        -2\Re\lambda_5)
       +\sin\beta[\sin^2\beta\Re\lambda_6
          +\cos^2\beta\Re(\lambda_7-2\lambda_6)]\},
 \\
a_{222}=\fr{1}{2}(\sin\beta\lambda_2+\cos\beta\Re\lambda_7),\quad
a_{223}=-\fr{1}{2}[\cos\beta\sin\beta\Im\lambda_5
           +(1+2\sin^2\beta)\Im\lambda_7],
 \\
a_{233}=\fr{1}{2}
       \{\sin\beta(\cos^2\beta\lambda_2+\sin^2\beta\Re\lambda_{345}
            -2\Re\lambda_5)
               +\cos\beta[\cos^2\beta\Re\lambda_7+\sin^2\beta\Re(\lambda_6-2\lambda_7)
           ]\},
\\
a_{333}=\fr{1}{2}(\cos\beta\sin\beta\Im\lambda_5-\sin^2\beta\Im\lambda_6
            -\cos^2\beta\Im\lambda_7).
 \label{Eq:trilin-coupl-a_mno}
\end{array}\ee

\subsubsection*{Quartic  couplings}
The quartic couplings involving two charged and two neutral Higgs
fields are given by Eq.~(\ref{Eq:charged-coupl}), where:
\begin{align}
b_{11}&=\sin^2\beta\lambda_1+\cos^2\beta\lambda_3
-2\cos\beta\sin\beta\Re\lambda_6, \nonumber\\
b_{12}&=-2[\cos\beta\sin\beta(\lambda_4+\Re\lambda_5)
-\sin^2\beta\Re\lambda_6-\cos^2\beta\Re\lambda_7], \nonumber\\
b_{13}&=2\cos\beta[\cos\beta\sin\beta\Im\lambda_5
-\sin^2\beta\Im\lambda_6-\cos^2\beta\Im\lambda_7], \nonumber\\
b_{22}&=\cos^2\beta\lambda_2+\sin^2\beta\lambda_3
-2\cos\beta\sin\beta\Re\lambda_7, \nonumber\\
b_{23}&=2\sin\beta[\cos\beta\sin\beta\Im\lambda_5
-\sin^2\beta\Im\lambda_6-\cos^2\beta\Im\lambda_7], \\
b_{33}&=\sin^4\beta\lambda_1+\cos^4\beta\lambda_2
+2\cos^2\beta\sin^2\beta\Re\lambda_{345}
-4\cos\beta\sin\beta(\sin^2\beta\Re\lambda_6+\cos^2\beta\Re\lambda_7).\nonumber
\end{align}

The quadrilinear couplings among four neutral Higgs fields are
given by Eq.~(\ref{Eq:neutral-3-4-coupl}), where:
 \bear{c}
a_{1111}=\fr{1}{8}\lambda_1,\;\,
a_{1112}=\fr{1}{2}\Re\lambda_6,\;\,
a_{1113}=-\fr{1}{2}\cos\beta\Im\lambda_6,\;\,
a_{1122}=\fr{1}{4}\Re\lambda_{345},\\[2mm]
a_{1123}=-\fr{1}{2}(\cos\beta\Im\lambda_5+\sin\beta\Im\lambda_6),\;\,
a_{1223}=-\fr{1}{2}[\sin\beta\Im\lambda_5+\cos\beta\Im\lambda_7],\\
a_{1133}=\fr{1}{4}[\sin^2\beta\lambda_1
+\cos^2\beta(\lambda_3+\lambda_4-\Re\lambda_5)
-2\cos\beta\sin\beta\Re\lambda_6],\\
a_{1222}=\fr{1}{2}\Re\lambda_7, \quad a_{2222}=\fr{1}{8}\lambda_2,
\quad a_{2223}=\fr{1}{2}\sin\beta\Im\lambda_7,
\\
a_{1233}=-\fr{1}{2}[-2\cos\beta\sin\beta\Re\lambda_5
+\sin^2\beta\Re\lambda_6+\cos^2\Im\lambda_7], \\
a_{1333}=\fr{1}{2}\cos\beta[\cos\beta\sin\beta\Im\lambda_5
-\sin^2\beta\lambda_6-\cos^2\beta\Im\lambda_7],\\
 a_{2233}=\fr{1}{4}[\cos^2\beta\lambda_2
+\sin^2\beta(\lambda_3+\lambda_4-\Re\lambda_5)
-2\cos\beta\sin\beta\Re\lambda_7],\\
a_{2333}=\fr{1}{2}\sin\beta[\cos\beta\sin\beta\Im\lambda_5
-\sin^2\beta\Im\lambda_6-\cos^2\Im\lambda_7],\\
a_{3333}=\fr{1}{8} [\sin^4\beta\lambda_1+\cos^4\beta\lambda_2
 +2\cos^2\beta\sin^2\beta\Re\lambda_{345}
-4\sin\beta\cos\beta(\sin^2\beta\Re\lambda_6+\cos^2\beta\Re\lambda_7)].
 \end{array}\label{Eq:quartic-coupl-a_mnop}
 \ee

\subsection{The CP-conserving, soft $Z_2$ violating case}


Below we  collect couplings for the CP-conserving, explicitly soft
$Z_2$ violating case,
$\lambda_6=\lambda_7=\Im\lambda_5=0$.\vspace{2mm}

\subsubsection{\bm Couplings in terms of $\lambda_i$, $\alpha$,
$\beta$}

First,  for completeness we present { well known in literature}
mentioned couplings  using our  potential.

\subsubsection*{Trilinear couplings}
For the CP-even Higgs bosons we have
\begin{eqnarray}
g_{hhh}&=& 3 v \bigl[-\cos\beta\sin^3\alpha\,\lambda_1
      +\sin\beta\cos^3\alpha\,\lambda_2
      -\fr{1}{2}\sin2\alpha \cos(\beta+\alpha)\lambda_{345}\bigr],
\nonumber \\
g_{Hhh}&=& v \bigl\{3\cos\beta\cos\alpha\sin^2\alpha\,\lambda_1
      +3\sin\beta\sin\alpha\cos^2\alpha\,\lambda_2 
   +[(1- 3\sin^2\alpha)\cos(\beta+\alpha)
-\sin\beta\sin\alpha]\lambda_{345}\bigr\}, \nonumber \\
g_{HHh}&=& v \bigl\{-3\cos\beta\sin\alpha\cos^2\alpha\,\lambda_1
       +3\sin\beta\cos\alpha\sin^2\alpha\,\lambda_2 
   +[\cos\beta\sin\alpha+
     (1-3\sin^2\alpha)\sin(\beta+\alpha)]\lambda_{345}\bigr\},
\nonumber \\
g_{HHH}&=&3v \bigl[\cos\beta\cos^3\alpha\,\lambda_1
     +\sin\beta\sin^3\alpha\,\lambda_2
     +\fr{1}{2}\sin2\alpha \sin(\beta+\alpha)\lambda_{345}\bigr].
\end{eqnarray}
For couplings involving the $CP$-odd $A$ we have
\begin{eqnarray}
g_{AAA}&=& \quad g_{Ahh}= \quad g_{AHH}= \quad
g_{AHh}= \quad g_{AH^+H^-}\;=\;0, \nonumber \\
g_{AAh}&=& v \bigl[-\cos\beta\sin^2\beta\sin\alpha\,\lambda_1
      +\sin\beta\cos^2\beta\cos\alpha\,\lambda_2 \nonumber\\
&& \quad
      +(\sin^3\beta\cos\alpha-\cos^3\beta\sin\alpha)\,\lambda_{345}
      -2\sin(\beta-\alpha)\,\lambda_5\bigr], \\
g_{AAH}&=& v \bigl[\cos\beta\sin^2\beta\cos\alpha\,\lambda_1
     +\sin\beta\cos^2\beta\sin\alpha\,\lambda_2 \nonumber\\
&& \quad
     +(\cos^3\beta\cos\alpha+\sin^3\beta\sin\alpha)\,\lambda_{345}
     -2\cos(\beta-\alpha)\,\lambda_5\bigr]. \nonumber
\end{eqnarray}

In the charged Higgs sector we have
\begin{align}
\fr{g_{hH^+H^-}}{v}
&=\fr{\sin2\beta}{2}[\sin\beta\sin\alpha\,\lambda_1
-\cos\beta\cos\alpha\,\lambda_2 +\cos(\beta+\alpha)\lambda_{345}]
-\sin(\beta-\alpha)\lambda_3 \nonumber\\
\fr{g_{HH^+H^-}}{v}
&=\fr{\sin2\beta}{2}[\sin\beta\cos\alpha\,\lambda_1
+\cos\beta\sin\alpha\,\lambda_2 -\sin(\beta+\alpha)\lambda_{345}]
+\cos(\beta-\alpha)\lambda_3\,. \label{chargebas}
\end{align}

\subsubsection*{Quartic couplings}
In the $CP$-even sector we have:
\begin{eqnarray}
g_{hhhh}&=&3 [\sin^4\alpha\,\lambda_1 + \cos^4\alpha\,\lambda_2
       +\fr{1}{2}\sin^2 2\alpha\,\lambda_{345}], \nonumber \\
g_{hhhH}&=&fr{3}{2} \sin2\alpha
       [-\sin^2\alpha\,\lambda_1 +\cos^2\alpha\,\lambda_2
       -\cos2\alpha\,\lambda_{345}], \nonumber \\
g_{hhHH}&=&\fr{3}{4}\sin^2 2\alpha(\lambda_1+\lambda_2)
      +(1-fr{3}{2}\sin^2 2\alpha)\lambda_{345}, \nonumber \\
g_{hHHH}&=&fr{3}{2} \sin2\alpha
       [-\cos^2\alpha\,\lambda_1+\sin^2\alpha\,\lambda_2
       +\cos2\alpha\,\lambda_{345}], \nonumber \\
g_{HHHH}&=&3[\cos^4\alpha\,\lambda_1 + \sin^4\alpha\,\lambda_2
       +\fr{1}{2}\sin^2 2\alpha\,\lambda_{345}].
\end{eqnarray}

Quartic couplings involving the $CP$-odd $A$:
\begin{eqnarray}
g_{hhhA}&=&g_{hhHA}=g_{hHHA}=g_{HHHA}=g_{hAAA}=g_{HAAA}=0, \nonumber \\
g_{hhAA}&=&\sin^2\beta\sin^2\alpha\,\lambda_1
             +\cos^2\beta\cos^2\alpha\,\lambda_2 \nonumber \\
&&\quad
 +(\cos^2\beta\sin^2\alpha+\sin^2\beta\cos^2\alpha)\lambda_{345}
  -\{1-\cos[2(\beta-\alpha)]\}\lambda_5, \nonumber \\
g_{hHAA}&=&\fr{1}{2}\sin2\alpha
[-\sin^2\beta\,\lambda_1+\cos^2\beta\,\lambda_2
 -\cos2\beta\,\lambda_{345}]
 -\sin[2(\beta-\alpha)]\lambda_5, \nonumber \\
g_{HHAA}&=&\sin^2\beta\cos^2\alpha\,\lambda_1
             +\cos^2\beta\sin^2\alpha\,\lambda_2 \nonumber \\
&&\quad
  +(\cos^2\beta\cos^2\alpha+\sin^2\beta\sin^2\alpha)\lambda_{345}
  -\{1+\cos[2(\beta-\alpha)]\}\lambda_5, \nonumber \\
g_{AAAA}&=&3[\sin^4\beta\,\lambda_1+\cos^4\beta\,\lambda_2
+\fr{1}{2}\sin^2 2\beta\,\lambda_{345}].
\end{eqnarray}

Quartic couplings involving the charged Higgs bosons:
\begin{eqnarray}
g_{hhH^+H^-}&=&\sin^2\beta\sin^2\alpha\,\lambda_1
                 +\cos^2\beta\cos^2\alpha\,\lambda_2
  +\fr{1}{2}\{1-\cos[2(\beta-\alpha)]\}\lambda_3 
  +\fr{1}{2}\sin2\beta\sin2\alpha\,\lambda_{345}, \nonumber \\
g_{hHH^+H^-}&=&\fr{1}{2}\sin2\alpha
(-\sin^2\beta\,\lambda_1+\cos^2\beta\,\lambda_2)
+\fr{1}{2}\sin[2(\beta-\alpha)]\lambda_3 
-\fr{1}{2}\sin2\beta\cos2\alpha\,\lambda_{345},
\nonumber \\
g_{HHH^+H^-}&=&\sin^2\beta\cos^2\alpha\,\lambda_1
                 +\cos^2\beta\sin^2\alpha\,\lambda_2
  +\fr{1}{2}\{1+\cos[2(\beta-\alpha)]\}\lambda_3 
 -\fr{1}{2}\sin2\beta\sin2\alpha\,\lambda_{345}, \nonumber \\
g_{hAH^+H^-}&=&g_{HAH^+H^-}=0, \nonumber \\
g_{AAH^+H^-}&=&\sin^4\beta\,\lambda_1+\cos^4\beta\,\lambda_2
+\fr{1}{2}\sin^2 2\beta\,\lambda_{345},
\nonumber \\
g_{H^+H^-H^+H^-}&=&2[\sin^4\beta\,\lambda_1+\cos^4\beta\,\lambda_2
+\fr{1}{2}\sin^2 2\beta\,\lambda_{345}].
\end{eqnarray}

\subsubsection{Trilinear couplings in terms of masses}

It is useful to express  parameters  $\lambda_i$ via Higgs boson
masses and mixing angles with the aid of eqs. (3.5), (3.12). We
get
 \bear{c}
 \label{Eq:lambdas}
\lambda_1 =\dfrac{1}{\cos^2\beta} \left[\fr{\cos^2\alpha\, M_H^2 +
\sin^2\alpha\, M_h^2}{v^2} - \nu \,\sin^2\beta\right] \,,
\\[5mm]
\lambda_2  =\dfrac{1}{\sin^2\beta} \left[\fr{\sin^2\alpha\, M_H^2
+ \cos^2\alpha\, M_h^2}{v^2} - \nu\,\cos^2\beta\right]\,,
 \\[5mm]
\lambda_{345}=\fr{\sin2\alpha}{\sin2\beta}\, \fr{M_H^2-M_h^2}{v^2}
+\nu,\quad
\lambda_4 =\fr{M_A^2-2M_{H^\pm}^2}{v^2}+\nu  , \quad \lambda_5
=-\fr{M_A^2}{v^2}+\nu.
\end{array}\ee
Now one can express triple Higgs couplings via masses $\beta$ and
$\alpha$ -- this way a dependence on the parameter $\nu $ emerges.

 For CP-even Higgs bosons
 \bear{c}
g_{hhh}=\dfrac{3}{v\sin2\beta}
\bigl[(\cos\beta\cos^3\alpha-\sin\beta\sin^3\alpha)
M_h^2-\cos^2(\beta-\alpha)\cos(\beta+\alpha)\nu v^2\bigr],
 \\[3mm]
g_{Hhh}=\dfrac{1}{2v\sin2\beta}
\cos(\beta-\alpha)\bigl[\sin2\alpha(2M_h^2+M_H^2)
+(\sin2\beta-3\sin2\alpha)\nu v^2\bigr],
 \\[3mm]
g_{HHh}=\dfrac{1}{2v\sin2\beta }
\sin(\beta-\alpha)\bigl[-\sin2\alpha(M_h^2+2M_H^2)
 +(\sin2\beta+3\sin2\alpha)\nu v^2\bigr],  \\[3mm]
g_{HHH}=\dfrac{3}{v\sin2\beta}
\bigl[(\sin\beta\cos^3\alpha+\cos\beta\sin^3\alpha)M_H^2
-\sin^2(\beta-\alpha)\sin(\beta+\alpha)\nu v^2\bigr],
\end{array}\ee
For interactions with $A$
 \bear{c}
g_{hAA}=\dfrac{1}{v}
\left[(2M_A^2-M_h^2)\sin(\beta-\alpha)+(M_h^2-\nu
v^2)\dfrac{\cos({\alpha+\beta})}{\sin\beta \cos\beta}\right]
 \,,\\[3mm]
g_{HAA}=\dfrac{1}{v}\left[(2M_A^2-M_H^2)\cos(\beta-\alpha)+(M_H^2-\nu
v^2)\dfrac{\cos(\alpha+\beta)}{\sin\beta \cos\beta}\right] \, .
\end{array}
\ee For interactions with $H^\pm$
 \bear{cl}
g_{h H^+H^-} &=\;\dfrac{1}{v}\,
\left[\left(2M_{H^\pm}^2-M_h^2\right)\sin(\beta-\alpha)+
 \dfrac{(M_h^2-\nu v^2)\cos(\beta+\alpha)}
{\sin\beta \cos\beta}\right]\\
 &=\;\dfrac{1}{v}\,
\left[\left(2M_{H^\pm}^2+M_h^2-2\nu v^2\right)\sin(\beta-\alpha)+
 2(M_h^2-\nu v^2)\cos(\beta-\alpha)\cot2\beta\right]
\,, \\
g_{H H^+H^-} &=\;\dfrac{1}{v}\,
\left[\left(2M_{H^\pm}^2-M_H^2\right)\cos(\beta-\alpha)
+\fr{(M_H^2-\nu v^2)\sin(\beta+\alpha)} {\sin\beta
\cos\beta}\right]\\
&=\;\dfrac{1}{v}\, \left[\left(2M_{H^\pm}^2+M_H^2-2\nu v^2
\right)\cos(\beta-\alpha)-
 2(M_H^2-\nu v^2)\sin(\beta-\alpha)\cot2\beta\right]
\,.\end{array}\label{b2d2} \ee

\subsubsection{Trilinear couplings  in terms of masses and relative
couplings in Model II}

For the Model II for the interaction with fermions we find
(denoting by
$\phi$ either $h$ or $H$)
 \bear{c}
g_{\phi\phi\phi}=\dfrac{3}{2v}\left[\left(
\chi_u^\phi+\chi_d^\phi-\chi_V^\phi\chi_u^\phi\chi_d^\phi \right)
(M_\phi^2-\nu v^2) +\chi_V^\phi\nu v^2\right]\,,\\[3mm]
g_{\phi_1\phi_2\phi_2}=-\dfrac{1}{2 v}\,\chi_V^{\phi_1}
\bigl[\chi_u^{\phi_2}\chi_d^{\phi_2}(2M_{\phi_2}^2+M_{\phi_1}^2-3\nu
v^2) -\nu v^2\bigr]\quad (\phi_1\neq \phi_2)\,,\\[3mm]
 g_{\phi AA}=\dfrac{1}{v}\left[(2M_A^2-M_\phi^2)\chi_V^\phi+
 (M_\phi^2-\nu v^2)(\chi_u^\phi+\chi_d^\phi)\right]\,,\\[3mm]
 g_{A\phi_1\phi_2}=g_{AAA}=g_{AH^+H^-}=0\,,\\[2mm]
g_{\phi H^+H^-} = \dfrac{1}{v}
\left[\left(2M_{H^\pm}^2-M_\phi^2\right)\chi^\phi_V +(M_\phi^2-\nu
v^2) (\chi^\phi_u+\chi^\phi_d)\right]\,.
\end{array} \label{triplII}\ee
\end{widetext}

\end{document}